\documentclass[aps,prd,twocolumn,showpacs,superscriptaddress,nofootinbib,preprintnumbers]{revtex4-2}

\usepackage[pdftex,colorlinks]{hyperref}
\usepackage[dvipsnames]{xcolor}
\usepackage{minted}
\hypersetup{
    colorlinks=true,
    linkcolor=NavyBlue,
    citecolor=NavyBlue,
    filecolor=NavyBlue,
    urlcolor=NavyBlue,
    }
\usepackage[pdftex]{graphicx}
\usepackage{bm}
\usepackage{cancel}
\usepackage{here}
\usepackage{comment,braket}
\usepackage{epsf}
\usepackage{amsmath}
\usepackage{graphics}
\usepackage{amsfonts}
\usepackage{amssymb}
\usepackage{latexsym}
\usepackage{color}
\usepackage{natbib}
\usepackage[normalem]{ulem}
\usepackage{orcidlink}
\usepackage{cleveref}
\usepackage{physics}
\usepackage{fontawesome5}
\DefineVerbatimEnvironment{commandbox}{Verbatim}{
frame=single,
framesep=4pt,
framerule=0.4pt,
fontsize=\small,
breaklines=true,
breakanywhere=true
}
\crefname{section}{Sec.}{Sec.}
\crefformat{section}{Sec.~#2#1#3}
\crefmultiformat{section}{Sec.~#2#1#3}{-#2#1#3}{, #2#1#3}{-#2#1#3}
\allowdisplaybreaks

\begin{document}

\preprint{\tt FERMILAB-PUB-26-0408-T}

\title{\texttt{LeWRON}: Agentic Analysis of Electroweak Phase Transitions}

\author{Isaac R. Wang\,\orcidlink{0000-0003-0789-218X}}
\email{isaacw@fnal.gov}
\affiliation{Theory Division, Fermi National Accelerator Laboratory, Batavia, IL, USA}

\begin{abstract}
The electroweak phase transition (EWPT) is a central topic in particle physics and cosmology, connecting collider phenomenology, baryogenesis, and gravitational-wave observatories.
Its analysis requires a technically demanding, convention-sensitive, and model-dependent pipeline, from constructing the finite-temperature effective potential to tracking thermal histories, computing bubble nucleation rates, and predicting gravitational-wave spectra.
We present \texttt{LeWRON} (\textbf{L}earning \textbf{E}lectro\textbf{W}eak phase t\textbf{R}ansiti\textbf{ON}), an agentic framework that orchestrates this pipeline starting from an input Lagrangian.
\texttt{LeWRON} combines audited toolbox construction with an Explorer module that uses the generated model-specific code for further analysis, including scans and plots.
Intermediate analytic outputs are checked by auditor agents and stored as structured artifacts, enabling reproducible human inspection and downstream use through both a command-line interface and a public \verb|Python| API.
The framework supports a reproduction mode, which infers conventions from the literature and reproduces published results, and a discovery mode, which guides users through structured checkpoints for new models.
We demonstrate \texttt{LeWRON} across representative beyond-the-Standard-Model scenarios and release the code on \href{https://github.com/quarkquartet/LeWRON}{\faGithub\ GitHub}.
\end{abstract}

\bigskip
\maketitle

{\bf Introduction.}
As the last discovered building block of the Standard Model (SM), the Higgs boson has been studied extensively over the past decade.
Nevertheless, its properties remain far from fully understood.
For example, current Higgs-coupling measurements still allow deviations at the few-percent to ten-percent level~\cite{ATLAS:2022vkf,CMS:2022dwd,ATLAS:2025qxq,CMS:2026nce}, and the Higgs trilinear self-coupling is constrained only at order-one precision~\cite{CMS:2026nce,CMS:2024awa,CMS:2026nuu,ATLAS:2025hhd}, and the exotic Higgs decay branching fractions are still allowed at the $10^{-2}$ level~\cite{ATLAS:2025qyn}.

These unresolved properties make the history of electroweak symmetry breaking an open question.
If electroweak symmetry was restored in the early Universe, its subsequent breaking proceeded through an electroweak phase transition (EWPT).
In the SM this transition is a smooth crossover~\cite{Kajantie:1993ag,Farakos:1994kj,Jansen:1995yg,Kajantie:1995kf,Rummukainen:1996sx,Kajantie:1996mn,Gurtler:1997hr,Csikor:1998eu,Laine:1998vn,Laine:1998qk,Rummukainen:1998as,Fodor:1999at}.
In many simple beyond-Standard-Model (BSM) extensions still allowed by current Higgs measurements, hwoever, it can become first order~\cite{Anderson:1991zb,Pietroni:1992in,Espinosa:1993bs,McDonald:1993ey,Choi:1993cv,Joyce:1994bi,Cline:1995dg,Basler:2016obg,Basler:2017uxn,Jeong:2018jqe,Jeong:2018ucz,Grzadkowski:2018nbc,Basler:2019iuu,Carena:2019une,Chao:2019smr,Xie:2020wzn,Xie:2020bkl,Fernandez-Martinez:2020szk,Basler:2021kgq,Biermann:2022meg,Anisha:2022hgv,Anisha:2023vvu,Fernandez-Martinez:2022stj,Harigaya:2023bmp}.
Alternatively, electroweak symmetry may never have been restored at high temperature (see e.g. Ref.~\cite{Carena:2021onl,Chang:2022psj} for recent studies).
These possibilities make the electroweak thermal history a sensitive probe of the Higgs sector.
A strong first-order EWPT (SFOPT) can also provide the out-of-equilibrium dynamics required for baryogenesis~\cite{Kuzmin:1985mm,Shaposhnikov:1986jp,Farrar:1993hn,Farrar:1993sp}, and a sufficiently strong first-order transition may source a gravitational-wave signal~\cite{Kamionkowski:1993fg,Caprini:2007xq,Huber:2008hg,Hindmarsh:2013xza,Hindmarsh:2016lnk,Kosowsky:2002,Caprini:2006,Caprini:2009yp}.

\begin{figure*}[t!]
  \centering
  \includegraphics[width=0.8\linewidth]{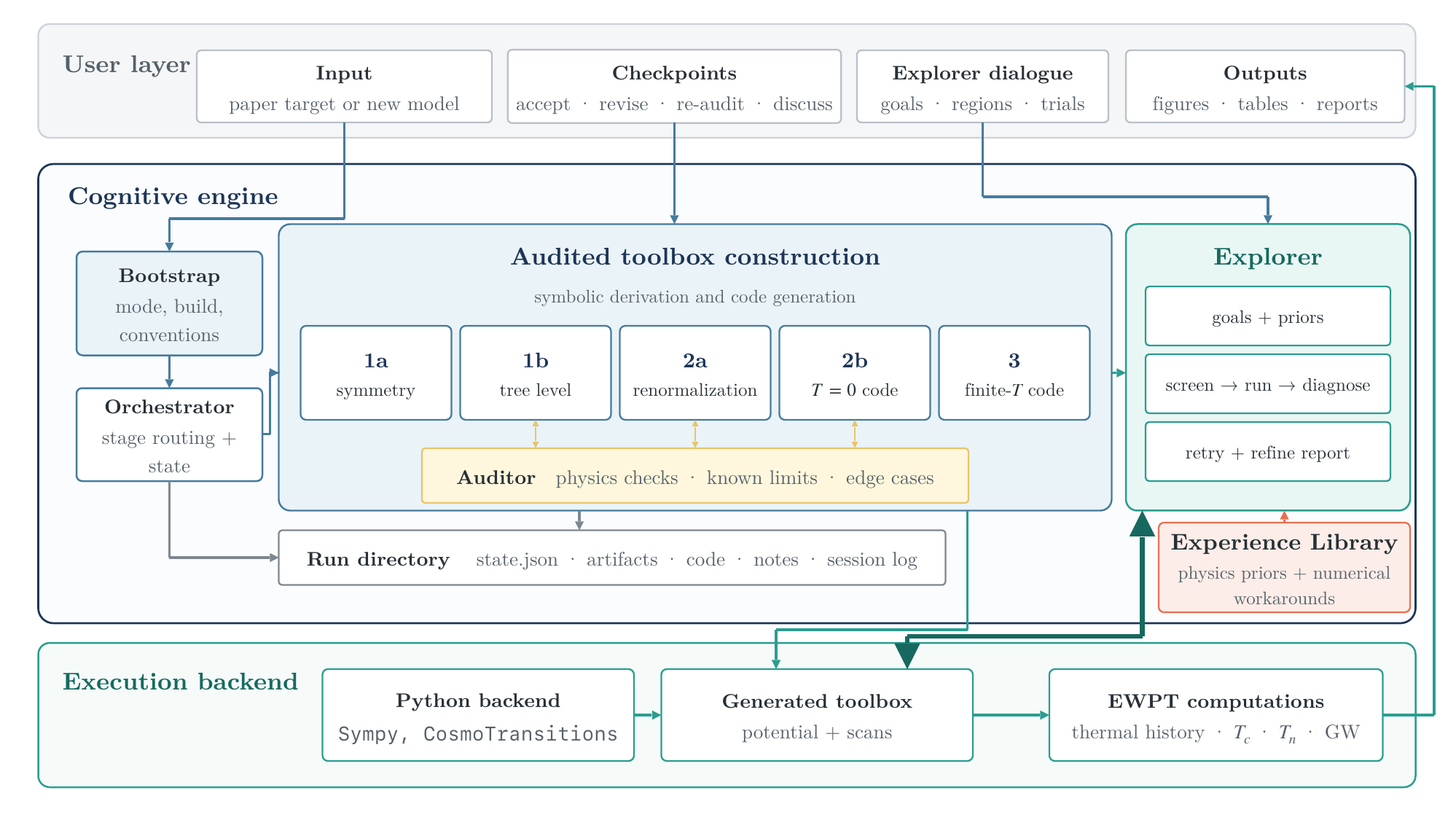}
  \vspace*{-0.2cm}
  \caption{Architecture of \texttt{LeWRON}.}
  \label{fig:architecture}
  \vspace*{-0.4cm}
\end{figure*}

Despite its broad physics reach, EWPT analysis remains technically demanding.
For each new BSM model, one must derive the background field potential, field-dependent mass spectra, loop corrections and renormalization, thermal corrections at a chosen resummation scheme, and a numerical implementation following the theory scheme.
The result is also sensitive to convention choices and implementation details.
Public libraries such as \texttt{CosmoTransitions}~\cite{Wainwright:2011kj} provide valuable helper functions, but practical applications still typically require dedicated coding and numerical troubleshooting.
This situation contrasts with collider phenomenology, where mature run-card-based tools organize the pipeline.

In this work, we develop \texttt{LeWRON} (\textbf{L}earning \textbf{E}lectro\textbf{W}eak phase t\textbf{R}ansiti\textbf{ON}), an agentic framework for EWPT analysis.
Starting from either a BSM Lagrangian or a target result in the literature, \texttt{LeWRON} constructs a model-specific toolbox for the effective potential and finite-temperature computations, then uses it to study phase histories, tunneling, and gravitational-wave observables.
Auditor modules examine the analytical results at first, while human checkpoints then allow the user to review conventions and physics correctness, request revisions, and preserve the full calculation record.

Recent hep-ph agents have shown that LLMs can coordinate collider workflows~\cite{Menzo:2025cim,Agrawal:2026lvg,Desai:2026nmx,Menzo:2026qrl,Qiu:2026iby,Bakshi:2025fgx,Plehn:2026gxv}.
this activity has also motivated dedicated collider-agent benchmarks such as \texttt{Collider-Bench}~\cite{Faroughy:2026dkj}.
\texttt{LeWRON} addresses a different setting: BSM particle-cosmology problems in which the analytic ingredients must be derived before numerical exploration can begin.
To our knowledge, it is the first domain-specialized agentic framework demonstrated on end-to-end EWPT analyses without assuming a mature model-specific pipeline\footnote{In concurrent work, \texttt{DarkAgents}~\cite{Lucente:2026kgh} presents an agentic system for dark first-order phase transitions using a fixed, human-written semi-analytic backend and the external \textsc{PTArcade} sampler.}.

{\bf Architecture.}
The central problem for \texttt{LeWRON} is the reliability and reproducibility of agent work, especially for analytical derivations.
\texttt{LeWRON} resolves this tension through the following design principles:
\begin{itemize}
\item {\it Role-separated model calls.}
Mechanical tasks with sharply specified targets, such as symbolic derivation, are assigned to a coding model with extended thinking disabled and \verb|temperature=0|, while diagnosis, exploration, numerical implementation, and auditing are assigned to a reasoning model.
This reduces stochastic variation in convention-following stages while reserving stronger reasoning for steps that require physics judgment.
For schema-bound derivations, lower-thinking deterministic calls often proved more reliable at following stage instructions than stronger reasoning configurations.

\item {\it Computer-algebra-assisted symbolic derivations.}
We divide the most involved derivations into smaller substeps and delegate symbolic manipulations to a computer algebra system (CAS).
In particular, stage~1b is decomposed into 10 substeps, each with its own auditor, ensuring that the generated code is checked against the corresponding derivation.

\item {\it Audited intermediate artifacts.}
Each analytic or semi-analytic stage produces structured artifacts, such as background directions, potentials, tadpoles, mass matrices, counterterms, and thermal masses.
Auditor agents review these artifacts before the pipeline advances to the next stage or substep; failed checks trigger regeneration using the auditor's feedback.

\begin{figure*}[t!]
    \centering
    \includegraphics[width=0.376\linewidth]{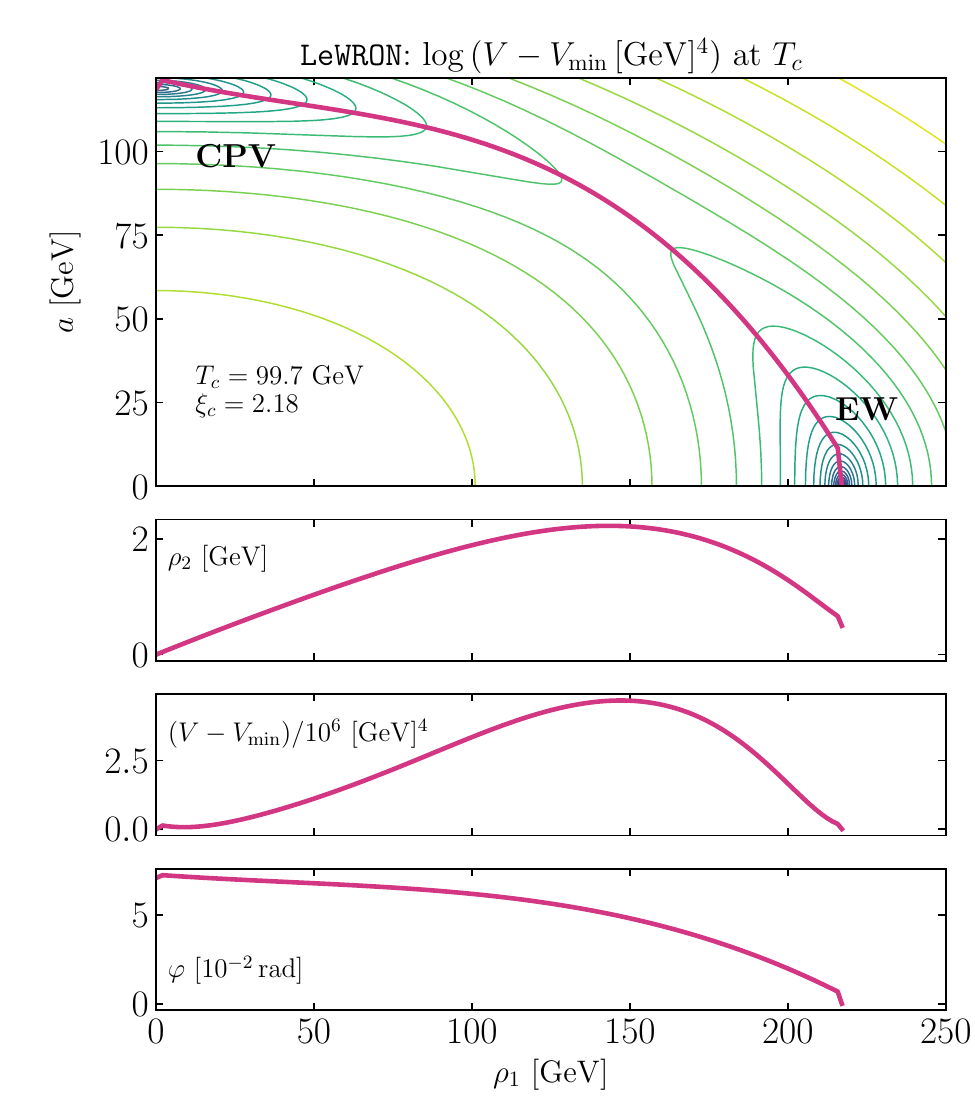}
    \includegraphics[width=0.35\linewidth]{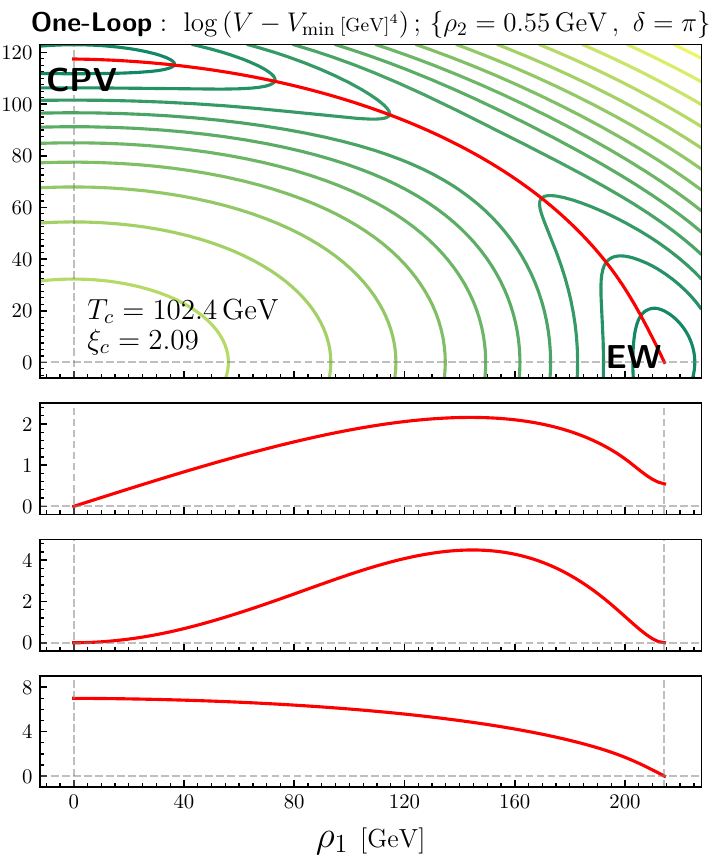}
    \caption{Reproduction of the right panel of Fig.~1 in Ref.~\cite{Gent:2025csq}. Left: result from \texttt{LeWRON}. Right: result copied from Ref.~\cite{Gent:2025csq}.}
    \label{fig:2512}
\end{figure*}

\item {\it Human-controlled physics review.}
At checkpoints of discovery mode, the user can inspect the current artifacts, ask questions, correct assumptions, provide model-specific guidance, request revision, or rerun the stage with additional instructions.
In reproduction mode, these choices are collected into a single reconstruction plan.

\end{itemize}
Together, these mechanisms make \texttt{LeWRON} a controlled human-agent workflow rather than a one-shot code generator: auditor agents check analytic consistency, while human physicists fix assumptions and make final checks on the results.

Fig.~\ref{fig:architecture} summarizes the architecture of \texttt{LeWRON}.
The workflow has two components.
The first is an audited toolbox-construction pipeline that builds the model-specific machinery for EWPT calculations.
At the start of this pipeline, a bootstrap module parses the input task, checks that it lies within scope, and initializes the run directory to contain the generated code, reports, logs, and structured artifacts.
The second component is the Explorer module, which uses the Claude Code SDK together with a customized experience library\footnote{The experience library is user-customizable, and could be easily installed as Claude Code skills for users who prefer to run the Explorer directly in Claude Code.} to run downstream tasks with the generated toolbox, such as computing phase-transition quantities, producing plots, or scanning parameter space.
This design separates model-specific tool construction, where reproducibility is prioritized, from downstream phenomenological exploration, where flexibility is the first principle.

More specifically, the toolbox-construction pipeline consists of three stages.
The physics conventions and stage-level derivations used in the calculations are described in Sec.~1 of the Supplemental Material.
Stage~1 identifies the background field directions and derives the tree-level physics, including the scalar potential in terms of the background fields, mass spectrum, Goldstone modes identification, vacuum conditions, bounded-from-below constraints, non-tachyonic conditions, global vacuum expectation value (vev) condition, and a parametrization plan using symbolic \verb|SymPy| routines.
Stage~2 constructs the zero-temperature one-loop effective potential, fixes the renormalization conditions, and implements the model in a \verb|Python| script following the structure of \verb|CosmoTransitions|~\cite{Wainwright:2011kj}.
Stage~3 adds the finite-temperature potential, including thermal masses and a user-selected daisy-resummation prescription, with Parwani resummation~\cite{Parwani:1991gq} used by default.
Each stage writes both machine-readable \verb|Pydantic| artifacts and human-readable Markdown notes.

{\bf Validation examples.}
We validate \texttt{LeWRON} with examples in both reproduction and discovery modes.
Reproduction examples test whether \texttt{LeWRON} can infer literature conventions, generate the corresponding analytic and numerical implementation, and recover published results.
Reproduction mode is not a truth certificate for the target paper: if \texttt{LeWRON} finds inconsistent formulae or conventions, it records them in the report, while the numerical computations in the following stages still respect the paper's formulae.
If the numerical result cannot be consistently reproduced, the Explorer reports the discrepancy to the user instead of counting the run as successful.
Discovery examples test the same machinery for open-ended tasks where, after the user provides additional input at checkpoints, \texttt{LeWRON} constructs the model-specific toolbox, then it executes user-specific input tasks in the Explorer.
For each example, we provide the input prompt, generated code, structured artifacts, notes, and final outputs in the \texttt{examples/} directory of the public \href{https://github.com/quarkquartet/LeWRON}{GitHub repository}.

\begin{table}[t]
\centering
\scriptsize
\setlength{\tabcolsep}{2.0pt}
\renewcommand{\arraystretch}{1.12}
\begin{tabular*}{\columnwidth}{@{\extracolsep{\fill}}ccc c cccc}
\hline
$m_s$ & $\lambda_{hs}$ & $w_0$ & Source
& $T_n$ & $T_c$ & $v_n/T_n$ & $r$ \\
\hline
$63$ & $0.9$ & $130$ & \texttt{LeWRON}
& $104.15$ & $105.55$ & $1.98$ & $1.02$ \\
     &       &       & Ref.~\cite{Friedlander:2020tnq}
& $103.59$ & $104.87$ & $2.02$ & $1.02$ \\
\hline
$81$ & $1.0$ & $110$ & \texttt{LeWRON}
& $124.85$ & $126.46$ & $1.33$ & $1.05$ \\
     &       &       & Ref.~\cite{Friedlander:2020tnq}
& $124.30$ & $125.43$ & $1.40$ & $1.03$ \\
\hline
$66$ & $0.3$ & $160$ & \texttt{LeWRON}
& $130.65$ & $132.63$ & $1.24$ & $1.05$ \\
     &       &       & Ref.~\cite{Friedlander:2020tnq}
& $130.53$ & $132.68$ & $1.28$ & $1.05$ \\
\hline
$105$ & $0.8$ & $110$ & \texttt{LeWRON}
& $131.57$ & $135.09$ & $1.18$ & $1.11$ \\
      &       &       & Ref.~\cite{Friedlander:2020tnq}
& $130.65$ & $134.46$ & $1.24$ & $1.11$ \\
\hline
\end{tabular*}
\caption{
Comparison of \texttt{LeWRON} results with benchmark points in Table 1 of Ref.~\cite{Friedlander:2020tnq}.
All dimensionful quantities are in unit GeV.
The wall velocity \(v_w\) is omitted.
}
\label{tab:singlet_reproduction}
\end{table}

\begin{figure*}[t!]
    \centering
    \includegraphics[width=0.31\linewidth]{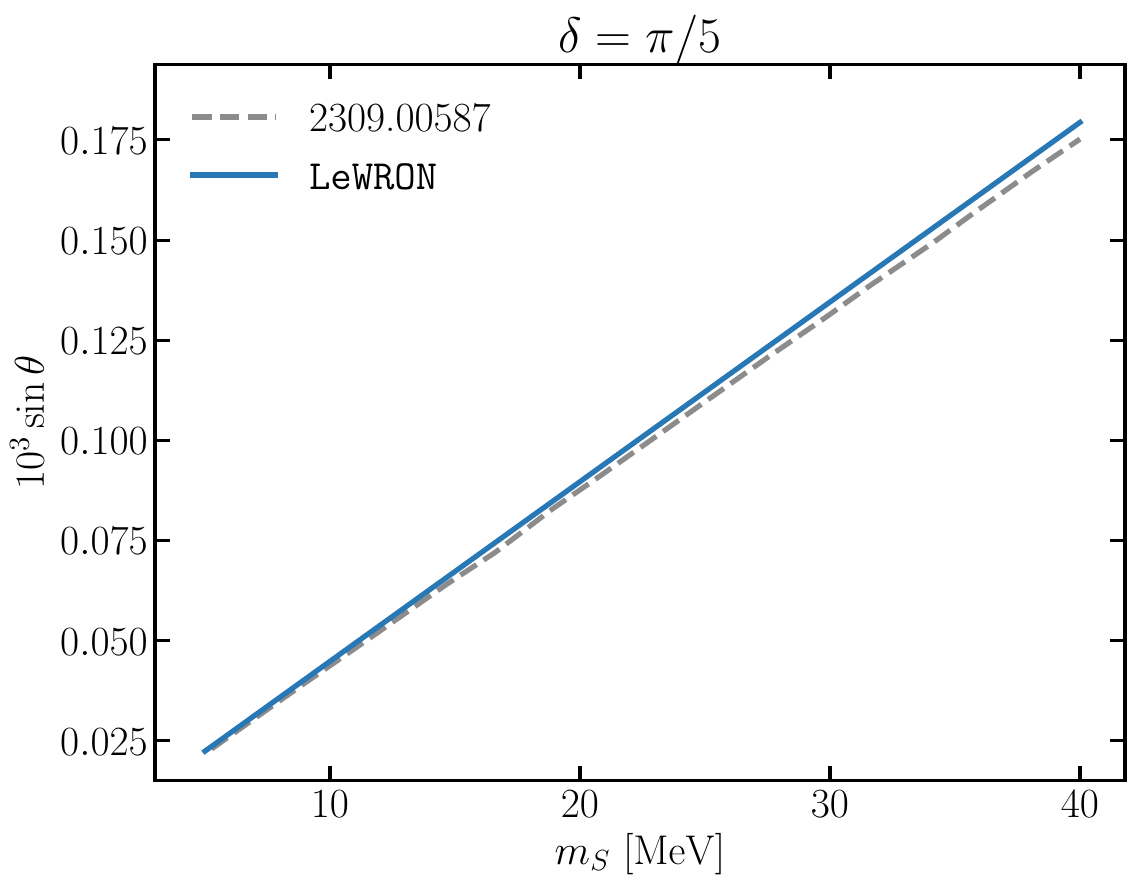}
    \includegraphics[width=0.68\linewidth]{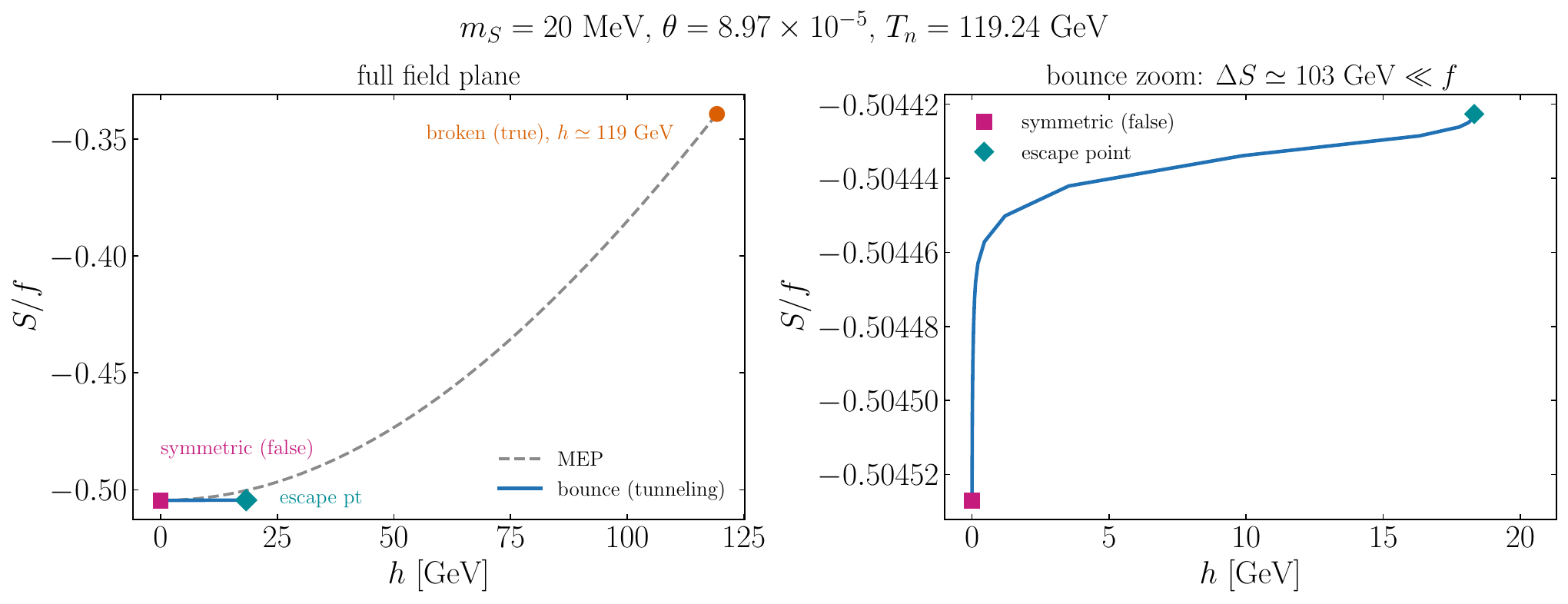}
    \caption{Tunneling action and parameter space of the ALP-Higgs model. Left: parameter space for a MeV-scale ALP, using the parameters of the middle-lower panel of Fig.~6 in Ref.~\cite{Harigaya:2023bmp} with $c=3$ (see also Sec.~S.3 of the Supplemental Material). The horizontal axis is the ALP mass, and the vertical axis is the ALP-Higgs mixing angle.
    The blue curve is the result from \texttt{LeWRON}, while the gray dashed line is digitized from the original work.
    Middle and right: tunneling path at nucleation for a parameter point chosen on the boundary in the left panel. The right panel zooms in on the middle panel. The blue solid line shows the solved tunneling path at the nucleation temperature.
    The gray dashed line shows the path that minimizes the effective potential along the tunneling direction, i.e., the minimal-energy path (MEP).}
    \label{fig:alp}
\end{figure*}

\textit{Reproduction mode} We use two examples to demonstrate the reproduction mode: Table~1 of Ref.~\cite{Friedlander:2020tnq} and Fig.~1 of Ref.~\cite{Gent:2025csq}.

Table~\ref{tab:singlet_reproduction} compares the output of \texttt{LeWRON} with Ref.~\cite{Friedlander:2020tnq}, which revisits the well-studied $\mathbb{Z}_2$-symmetric singlet scalar extension to investigate the bubble-wall velocity and thickness, with the nucleation temperature computed as a prerequisite.
This example shows that the agent can complete the standard computation from the Lagrangian to the nucleation temperature.
All results are in good agreement with an error smaller than 1\%.
The remaining discrepancy cannot be resolved from the information provided in the paper, as the exact input values of parameters, such as the gauge and Yukawa couplings, are lacking.

Fig.~\ref{fig:2512} compares the \texttt{LeWRON} result with the right panel of Fig.~1 in Ref.~\cite{Gent:2025csq}, which studies the potential at $T_c$ in the 2HDM+$a$ model.
The main technical difficulty is the rotation between the Higgs basis and the $Z_2$ basis.
For 2HDM-like models, \texttt{LeWRON} is equipped with dedicated knowledge and code templates for translating between common bases.
In reproduction mode, the basis choice of the target paper is respected.
The computed critical temperature differs from Ref.~\cite{Gent:2025csq} by about (3\%); no further source of this residual discrepancy was identified, which may originate from implementation-level numerical details, for instance in the treatment of thermal functions, minimization tolerances, or interpolation procedures.

\textit{Discovery mode} We use discovery mode to test whether the agent can handle a complete analysis of models.
We again consider two examples: the ALP-Higgs model~\cite{Jeong:2018jqe,Jeong:2018ucz,Harigaya:2023bmp}, and the 2HDM+$S$ model~\cite{Biermann:2022meg}.
In the discovery examples, the user refines target observables and final visualizations through  Explorer conversations.

Fig.~\ref{fig:alp} shows the result of the ALP-Higgs model.
\texttt{LeWRON} addresses two limitations in the previous literature.
First, the renormalization prescription used in Ref.~\cite{Harigaya:2023bmp} fixes all input parameters, while missing the additional required ALP-field tadpole term to preserve the ALP vev, leaving a residual inconsistency in the one-loop matching conditions, although numerically this correction has only a minor impact.
Second, the computation of the tunneling action for a MeV-scale ALP is numerically challenging.
Refs.~\cite{Jeong:2018jqe,Jeong:2018ucz} develop an analytic approximation to describe the tunneling path, while Ref.~\cite{Harigaya:2023bmp} extrapolates the relation between $T_n$ and $T_c$ near the SFOPT boundary.
\texttt{LeWRON} addresses both issues automatically by deriving a consistent renormalization scheme and implementing the tunneling-action computation.
As shown in the left panel, the SFOPT boundary derived by computing $v_n/T_n$ directly with the adjusted renormalization scheme agrees well with the extrapolation in Ref.~\cite{Harigaya:2023bmp}, indicating that the limitations of the previous treatment have only a minor impact on this boundary.
The tunneling path shown in the middle and right panels has the form expected from the analytic approximation in Refs.~\cite{Jeong:2018ucz,Jeong:2018jqe}.

\begin{figure}[t!]
    \centering
    \includegraphics[width=0.9\linewidth]{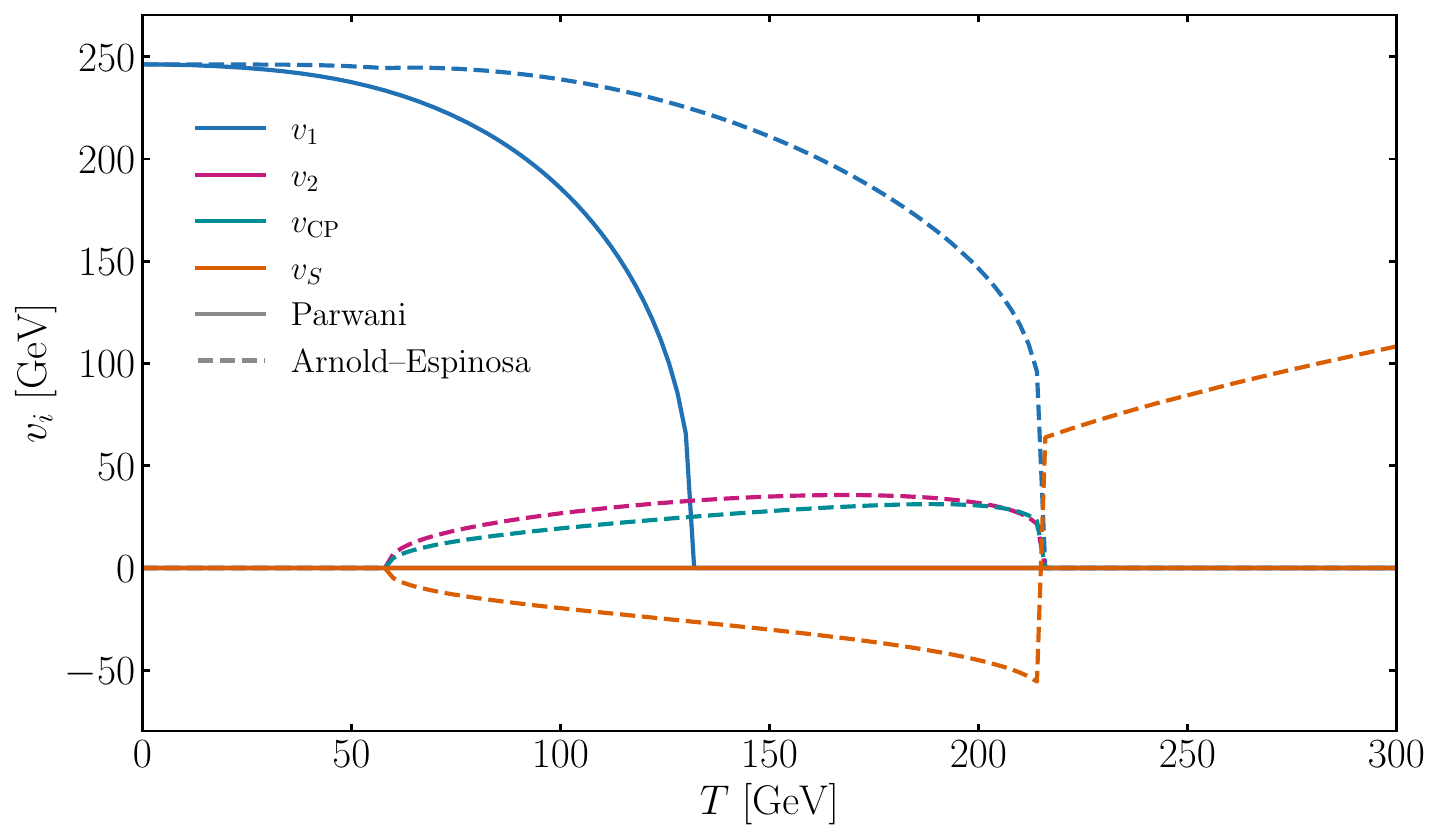}
    \caption{Thermal history of the 2HDM+$S$ model under different resummation schemes for the parameter point in Eq.~\eqref{eq:parameter}. The horizontal axis is the temperature, and the vertical axis shows the vevs of different degrees of freedom.}
    \label{fig:2hdm-S}
\end{figure}

Fig.~\ref{fig:2hdm-S} demonstrates the thermal history of the 2HDM+$S$ model, whose phase transition was first investigated in Ref.~\cite{Biermann:2022meg}.
We choose a benchmark point modified from Table~2(b) of that work
\begin{align}
    \label{eq:parameter}
&\lambda_1 = 0.258, ~ \lambda_2 = 3.660,~ \lambda_3 = -0.821, ~ \lambda_4 = 0.220 \nonumber \\
&\lambda_5 = -0.371., ~\lambda_6 = 6.0, ~ \lambda_7 = 4.0, ~\lambda_8 = 14.781,\nonumber \\
&m_{11}^2 = -7824\ \mathrm{GeV}^2, ~m_{22}^2 = 65258.809\ \mathrm{GeV}^2,\nonumber \\ 
&m_S^2 = 10^4\ \mathrm{GeV}^2,~ A_{\rm re} = 199.644\ \mathrm{GeV}, \nonumber \\
&A_{\rm im} = -233.318\ \mathrm{GeV}\,,
\end{align}
Our implementation differs from that setup in two ways.
First, the automatic procedure in \texttt{LeWRON} fixes 13 renormalization conditions for this model, corresponding to 5 quartic portal couplings, 5 physical masses, 1 Higgs vev, and 2 mixing angles.
By contrast, Ref.~\cite{Biermann:2022meg} takes all 9 field degrees of freedom into account, including Goldstone bosons.
We refer the reader to Sec.~3.2 of the Supplemental Material for more details.
Second, at the Stage~3 checkpoint we run the benchmark with both Parwani~\cite{Parwani:1991gq} and Arnold-Espinosa~\cite{Arnold:1992rz} resummation.
For each scheme, \texttt{LeWRON} constructs the corresponding finite-temperature implementation and reruns the Explorer for thermal-history search.
Comparing the resulting artifacts, we find qualitatively different thermal histories.
With Arnold-Espinosa resummation, the vev of the singlet $S$ is nonzero at high temperature before the Universe enters the electroweak vacuum.
This gives a first-order phase transition $(0,0,0, v_S) \to (v_1, v_2, v_\eta, v'_S)$, followed by a smooth crossover into the EW vev.
The Parwani resummation, on the other hand, gives a simple smooth crossover, $(0,0,0,0) \to (v,0,0,0)$.

{\bf Conclusion and outlook.}
We have presented \texttt{LeWRON}, an agentic framework for EWPT analysis that constructs model-specific analytic ingredients and numerical implementations, followed by a module that allows for interactive exploration.
The framework uses \verb|SymPy| to derive analytical formulas and pass the results through Pydantic artifacts, while using an independent LLM reasoning model as an auditor to check the results before passing to human checkpoints.

We validated the framework in four representative examples.
In reproduction mode, \texttt{LeWRON} reproduces benchmark nucleation temperatures in the $Z_2$-symmetric singlet model and reconstructs the finite-temperature potential structure of a 2HDM+$a$ benchmark with nontrivial basis translation.
In both examples, \texttt{LeWRON} shows excellent agreement with the input paper.
In discovery mode, starting from model descriptions rather than literature input, it explores the ALP-Higgs and 2HDM+$S$ examples, exhibiting a more consistent treatment and thus improving the results compared to the existing literature.

\texttt{LeWRON} is released as an open-source package with both a command-line interface and a public \verb|Python| API. The command-line interface supports full EWPT runs for physicists, while the \verb|Python| API allows users to incorporate individual \texttt{LeWRON} stages, modules, and substeps into external frameworks or model-behavior tests.

Several limitations remain for future work.
On the physics side, \texttt{LeWRON} does not yet compute the bubble-wall velocity, whose first-principles determination remains under active development; see, e.g., Refs.~\cite{Laurent:2022jrs,DeCurtis:2022hlx,
DeCurtis:2023hil,DeCurtis:2024hvh,Ai:2021kak,Ai:2023see,Ai:2024shx,
Ramsey-Musolf:2025jyk,Ai:2025bjw,Ai:2024btx,Carena:2025flp,
Krajewski:2024gma,Krajewski:2024zxg,Eriksson:2025owh} for recent progress.
The effective-potential calculation is also limited to the four-dimensional one-loop framework and does not yet include the three-dimensional effective-theory approach or higher loop orders.
We refer the readers to Ref.~\cite{Ginsparg:1980ef,Appelquist:1981vg,Braaten:1995cm,Kajantie:1995dw,Croon:2020cgk,Gould:2021dzl,Niemi:2021qvp,Schicho:2021gca,Ekstedt:2022bff} for original works and recent updates.

On the agent side, the main limitation is the tradeoff between reproducibility and flexibility.
The most constrained part of the workflow is stage~1b, where code helpers make symbolic derivations more reproducible but require detailed knowledge of supported model classes.
The restriction improves the reliability at the cost of model coverage and flexibility in accepting arbitrary user specifications.
This follows the same philosophy as the analytic-derivation component of \texttt{HEPTAPOD}~\cite{Menzo:2025cim}: reliability is improved by restricting the class of accepted inputs.
Consequently, the present version of \texttt{LeWRON} is limited to electroweak phase transitions driven by the SM Higgs sector.
For the same reason, electroweak-like transitions, such as the $SU(2)_R$ phase transition in left-right symmetric models~\cite{Brdar:2019fur,Harigaya:2022wzt}, dark phase transitions~\cite{Hall:2019ank,Hall:2019rld,Goncalves:2025uwh,Balan:2025uke,Costa:2025csj}, and models whose scalar quartics are fixed by another sector and therefore require a dedicated effective-potential and counterterm treatment, such as supersymmetric models, are left to future work.
The examples in this work were validated with Anthropic model APIs; systematic tests with other frontier-model providers are also left for future releases.

{\bf Acknowledgement} The author is grateful to Wenyue Hua for her encouragement, technical support, and many useful discussions throughout this project. The author thanks George Fleming and David Shih for insightful discussions and feedback on the manuscript.
The author also thanks Marcela Carena for her encouragement, mentorship, and support during the development of this work.

I.R.W.\, is supported by Fermi Forward Discovery Group, LLC under Contract No. 89243024CSC000002 with the U.S. Department of Energy, Office of Science, Office of High Energy Physics.
I.R.W.\, is also supported by DOE distinguished scientist fellowship grant FNAL 22-33.

\bibliographystyle{utphys3}
\bibliography{lewron}

\providecommand{\href}[2]{#2}\begingroup\raggedright\begin{thebibliography}{100}

\bibitem{ATLAS:2022vkf}
{\bfseries ATLAS} Collaboration, ``{A detailed map of Higgs boson interactions by the ATLAS experiment ten years after the discovery},'' \href{https://dx.doi.org/10.1038/s41586-022-04893-w}{{\em Nature} {\bfseries 607} no.~7917, (2022) 52--59}, \href{https://arxiv.org/abs/2207.00092}{{\ttfamily arXiv:2207.00092 [hep-ex]}}. Erratum: Nature 612, E24 (2022), doi:10.1038/s41586-022-05581-5.

\bibitem{CMS:2022dwd}
{\bfseries CMS} Collaboration, ``{A portrait of the Higgs boson by the CMS experiment ten years after the discovery},'' \href{https://dx.doi.org/10.1038/s41586-022-04892-x}{{\em Nature} {\bfseries 607} no.~7917, (2022) 60--68}, \href{https://arxiv.org/abs/2207.00043}{{\ttfamily arXiv:2207.00043 [hep-ex]}}. Erratum: doi:10.1038/s41586-023-06164-8.

\bibitem{ATLAS:2025qxq}
{\bfseries ATLAS} Collaboration, ``{Combined measurements of Higgs boson production and decay at $\sqrt{s} =$ 13 TeV using up to 140 fb$^{-1}$ of data collected by the ATLAS Experiment},''.

\bibitem{CMS:2026nce}
{\bfseries CMS} Collaboration, A.~Hayrapetyan {\em et~al.}, ``{Combined measurements and interpretations of Higgs boson production and decay in proton-proton collisions at $\sqrt{s}$ = 13 TeV},'' \href{https://arxiv.org/abs/2602.18611}{{\ttfamily arXiv:2602.18611 [hep-ex]}}.

\bibitem{CMS:2024awa}
{\bfseries CMS} Collaboration, A.~Hayrapetyan {\em et~al.}, ``{Constraints on the Higgs boson self-coupling from the combination of single and double Higgs boson production in proton-proton collisions at s=13TeV},'' \href{https://dx.doi.org/10.1016/j.physletb.2024.139210}{{\em Phys. Lett. B} {\bfseries 861} (2025) 139210}, \href{https://arxiv.org/abs/2407.13554}{{\ttfamily arXiv:2407.13554 [hep-ex]}}.

\bibitem{CMS:2026nuu}
{\bfseries CMS, ATLAS} Collaboration, G.~Aad {\em et~al.}, ``{Combination of ATLAS and CMS searches for Higgs boson pair production at $\sqrt{s} = 13$ TeV},'' \href{https://arxiv.org/abs/2602.23991}{{\ttfamily arXiv:2602.23991 [hep-ex]}}.

\bibitem{ATLAS:2025hhd}
{\bfseries ATLAS} Collaboration, G.~Aad {\em et~al.}, ``{Study of Higgs boson pair production in the $HH \rightarrow b \overline{b} \gamma\gamma$ final state with 308 fb$^{-1}$ of data collected at $\sqrt{s} = 13$ TeV and 13.6 TeV by the ATLAS experiment},'' \href{https://dx.doi.org/10.1016/j.physletb.2026.140280}{{\em Phys. Lett. B} {\bfseries 876} (2026) 140280}, \href{https://arxiv.org/abs/2507.03495}{{\ttfamily arXiv:2507.03495 [hep-ex]}}.

\bibitem{ATLAS:2025qyn}
{\bfseries ATLAS} Collaboration, G.~Aad {\em et~al.}, ``{Search for Higgs boson exotic decays into Lorentz-boosted light bosons in the four-{\ensuremath{\tau}} final state at s=13TeV with the ATLAS detector},'' \href{https://dx.doi.org/10.1016/j.physletb.2025.139843}{{\em Phys. Lett. B} {\bfseries 870} (2025) 139843}, \href{https://arxiv.org/abs/2503.05463}{{\ttfamily arXiv:2503.05463 [hep-ex]}}.

\bibitem{Kajantie:1993ag}
K.~Kajantie, K.~Rummukainen, and M.~E. Shaposhnikov, ``{A Lattice Monte Carlo study of the hot electroweak phase transition},'' \href{https://dx.doi.org/10.1016/0550-3213(93)90062-T}{{\em Nucl. Phys. B} {\bfseries 407} (1993) 356--372}, \href{https://arxiv.org/abs/hep-ph/9305345}{{\ttfamily arXiv:hep-ph/9305345}}.

\bibitem{Farakos:1994kj}
K.~Farakos, K.~Kajantie, K.~Rummukainen, and M.~E. Shaposhnikov, ``{The Electroweak phase transition at m(H) approximately = m(W)},'' \href{https://dx.doi.org/10.1016/0370-2693(94)90563-0}{{\em Phys. Lett. B} {\bfseries 336} (1994) 494--501}, \href{https://arxiv.org/abs/hep-ph/9405234}{{\ttfamily arXiv:hep-ph/9405234}}.

\bibitem{Jansen:1995yg}
K.~Jansen, ``Status of the {{Finite Temperature Electroweak Phase Transition}} on the {{Lattice}},'' \href{https://dx.doi.org/10.1016/0920-5632(96)00045-X}{{\em Nuclear Physics B - Proceedings Supplements} {\bfseries 47} (1996) 196--211}, \href{https://arxiv.org/abs/hep-lat/9509018}{{\ttfamily arXiv:hep-lat/9509018}}.

\bibitem{Kajantie:1995kf}
K.~Kajantie, M.~Laine, K.~Rummukainen, and M.~E. Shaposhnikov, ``{The Electroweak phase transition: A Nonperturbative analysis},'' \href{https://dx.doi.org/10.1016/0550-3213(96)00052-1}{{\em Nucl. Phys. B} {\bfseries 466} (1996) 189--258}, \href{https://arxiv.org/abs/hep-lat/9510020}{{\ttfamily arXiv:hep-lat/9510020}}.

\bibitem{Rummukainen:1996sx}
K.~Rummukainen, ``Finite {{T Electroweak Phase Transition}} on the {{Lattice}},'' \href{https://dx.doi.org/10.1016/S0920-5632(96)00597-X}{{\em Nuclear Physics B - Proceedings Supplements} {\bfseries 53} (1997) 30--42}, \href{https://arxiv.org/abs/hep-lat/9608079}{{\ttfamily arXiv:hep-lat/9608079}}.

\bibitem{Kajantie:1996mn}
K.~Kajantie, M.~Laine, K.~Rummukainen, and M.~E. Shaposhnikov, ``{Is there a~ hot electroweak phase transition at $m_H \gtrsim m_W$?},'' \href{https://dx.doi.org/10.1103/PhysRevLett.77.2887}{{\em Phys. Rev. Lett.} {\bfseries 77} (1996) 2887--2890}, \href{https://arxiv.org/abs/hep-ph/9605288}{{\ttfamily arXiv:hep-ph/9605288}}.

\bibitem{Gurtler:1997hr}
M.~Gurtler, E.-M. Ilgenfritz, and A.~Schiller, ``{Where the electroweak phase transition ends},'' \href{https://dx.doi.org/10.1103/PhysRevD.56.3888}{{\em Phys. Rev. D} {\bfseries 56} (1997) 3888--3895}, \href{https://arxiv.org/abs/hep-lat/9704013}{{\ttfamily arXiv:hep-lat/9704013}}.

\bibitem{Csikor:1998eu}
F.~Csikor, Z.~Fodor, and J.~Heitger, ``{Endpoint of the hot electroweak phase transition},'' \href{https://dx.doi.org/10.1103/PhysRevLett.82.21}{{\em Phys. Rev. Lett.} {\bfseries 82} (1999) 21--24}, \href{https://arxiv.org/abs/hep-ph/9809291}{{\ttfamily arXiv:hep-ph/9809291}}.

\bibitem{Laine:1998vn}
M.~Laine and K.~Rummukainen, ``{A Strong electroweak phase transition up to m(H) is about 105-GeV},'' \href{https://dx.doi.org/10.1103/PhysRevLett.80.5259}{{\em Phys. Rev. Lett.} {\bfseries 80} (1998) 5259--5262}, \href{https://arxiv.org/abs/hep-ph/9804255}{{\ttfamily arXiv:hep-ph/9804255}}.

\bibitem{Laine:1998qk}
M.~Laine and K.~Rummukainen, ``The {{MSSM Electroweak Phase Transition}} on the {{Lattice}},'' \href{https://dx.doi.org/10.1016/S0550-3213(98)00530-6}{{\em Nuclear Physics B} {\bfseries 535} (1998) 423--457}, \href{https://arxiv.org/abs/hep-lat/9804019}{{\ttfamily arXiv:hep-lat/9804019}}.

\bibitem{Rummukainen:1998as}
K.~Rummukainen, M.~Tsypin, K.~Kajantie, M.~Laine, and M.~E. Shaposhnikov, ``{The Universality class of the electroweak theory},'' \href{https://dx.doi.org/10.1016/S0550-3213(98)00494-5}{{\em Nucl. Phys. B} {\bfseries 532} (1998) 283--314}, \href{https://arxiv.org/abs/hep-lat/9805013}{{\ttfamily arXiv:hep-lat/9805013}}.

\bibitem{Fodor:1999at}
Z.~Fodor, ``Electroweak {{Phase Transitions}},'' \href{https://dx.doi.org/10.1016/S0920-5632(00)91603-7}{{\em Nuclear Physics B - Proceedings Supplements} {\bfseries 83--84} (2000) 121--125}, \href{https://arxiv.org/abs/hep-lat/9909162}{{\ttfamily arXiv:hep-lat/9909162}}.

\bibitem{Anderson:1991zb}
G.~W. Anderson and L.~J. Hall, ``{The Electroweak phase transition and baryogenesis},'' \href{https://dx.doi.org/10.1103/PhysRevD.45.2685}{{\em Phys. Rev. D} {\bfseries 45} (1992) 2685--2698}.

\bibitem{Pietroni:1992in}
M.~Pietroni, ``{The Electroweak phase transition in a nonminimal supersymmetric model},'' \href{https://dx.doi.org/10.1016/0550-3213(93)90635-3}{{\em Nucl. Phys. B} {\bfseries 402} (1993) 27--45}, \href{https://arxiv.org/abs/hep-ph/9207227}{{\ttfamily arXiv:hep-ph/9207227}}.

\bibitem{Espinosa:1993bs}
J.~R. Espinosa and M.~Quiros, ``{The Electroweak phase transition with a singlet},'' \href{https://dx.doi.org/10.1016/0370-2693(93)91111-Y}{{\em Phys. Lett. B} {\bfseries 305} (1993) 98--105}, \href{https://arxiv.org/abs/hep-ph/9301285}{{\ttfamily arXiv:hep-ph/9301285}}.

\bibitem{McDonald:1993ey}
J.~McDonald, ``{Electroweak baryogenesis and dark matter via a gauge singlet scalar},'' \href{https://dx.doi.org/10.1016/0370-2693(94)91229-7}{{\em Phys. Lett. B} {\bfseries 323} (1994) 339--346}.

\bibitem{Choi:1993cv}
J.~Choi and R.~R. Volkas, ``{Real Higgs singlet and the electroweak phase transition in the Standard Model},'' \href{https://dx.doi.org/10.1016/0370-2693(93)91013-D}{{\em Phys. Lett.} {\bfseries B317} (1993) 385--391}, \href{https://arxiv.org/abs/hep-ph/9308234}{{\ttfamily arXiv:hep-ph/9308234 [hep-ph]}}.

\bibitem{Joyce:1994bi}
M.~Joyce, T.~Prokopec, and N.~Turok, ``{Efficient electroweak baryogenesis from lepton transport},'' \href{https://dx.doi.org/10.1016/0370-2693(94)91377-3}{{\em Phys. Lett. B} {\bfseries 338} (1994) 269--275}, \href{https://arxiv.org/abs/hep-ph/9401352}{{\ttfamily arXiv:hep-ph/9401352}}.

\bibitem{Cline:1995dg}
J.~M. Cline, K.~Kainulainen, and A.~P. Vischer, ``{Dynamics of two Higgs doublet CP violation and baryogenesis at the electroweak phase transition},'' \href{https://dx.doi.org/10.1103/PhysRevD.54.2451}{{\em Phys. Rev. D} {\bfseries 54} (1996) 2451--2472}, \href{https://arxiv.org/abs/hep-ph/9506284}{{\ttfamily arXiv:hep-ph/9506284}}.

\bibitem{Basler:2016obg}
P.~Basler, M.~Krause, M.~Muhlleitner, J.~Wittbrodt, and A.~Wlotzka, ``{Strong First Order Electroweak Phase Transition in the CP-Conserving 2HDM Revisited},'' \href{https://dx.doi.org/10.1007/JHEP02(2017)121}{{\em JHEP} {\bfseries 02} (2017) 121}, \href{https://arxiv.org/abs/1612.04086}{{\ttfamily arXiv:1612.04086 [hep-ph]}}.

\bibitem{Basler:2017uxn}
P.~Basler, M.~M\"uhlleitner, and J.~Wittbrodt, ``{The CP-Violating 2HDM in Light of a Strong First Order Electroweak Phase Transition and Implications for Higgs Pair Production},'' \href{https://dx.doi.org/10.1007/JHEP03(2018)061}{{\em JHEP} {\bfseries 03} (2018) 061}, \href{https://arxiv.org/abs/1711.04097}{{\ttfamily arXiv:1711.04097 [hep-ph]}}.

\bibitem{Jeong:2018jqe}
K.~S. Jeong, T.~H. Jung, and C.~S. Shin, ``{Adiabatic electroweak baryogenesis driven by an axionlike particle},'' \href{https://dx.doi.org/10.1103/PhysRevD.101.035009}{{\em Phys. Rev. D} {\bfseries 101} no.~3, (2020) 035009}, \href{https://arxiv.org/abs/1811.03294}{{\ttfamily arXiv:1811.03294 [hep-ph]}}.

\bibitem{Jeong:2018ucz}
K.~S. Jeong, T.~H. Jung, and C.~S. Shin, ``{Axionic Electroweak Baryogenesis},'' \href{https://dx.doi.org/10.1016/j.physletb.2019.01.036}{{\em Phys. Lett. B} {\bfseries 790} (2019) 326--331}, \href{https://arxiv.org/abs/1806.02591}{{\ttfamily arXiv:1806.02591 [hep-ph]}}.

\bibitem{Grzadkowski:2018nbc}
B.~Grzadkowski and D.~Huang, ``{Spontaneous $CP$-Violating Electroweak Baryogenesis and Dark Matter from a Complex Singlet Scalar},'' \href{https://dx.doi.org/10.1007/JHEP08(2018)135}{{\em JHEP} {\bfseries 08} (2018) 135}, \href{https://arxiv.org/abs/1807.06987}{{\ttfamily arXiv:1807.06987 [hep-ph]}}.

\bibitem{Basler:2019iuu}
P.~Basler, M.~M\"uhlleitner, and J.~M\"uller, ``{Electroweak Phase Transition in Non-Minimal Higgs Sectors},'' \href{https://dx.doi.org/10.1007/JHEP05(2020)016}{{\em JHEP} {\bfseries 05} (2020) 016}, \href{https://arxiv.org/abs/1912.10477}{{\ttfamily arXiv:1912.10477 [hep-ph]}}.

\bibitem{Carena:2019une}
M.~Carena, Z.~Liu, and Y.~Wang, ``{Electroweak phase transition with spontaneous Z$_{2}$-breaking},'' \href{https://dx.doi.org/10.1007/JHEP08(2020)107}{{\em JHEP} {\bfseries 08} (2020) 107}, \href{https://arxiv.org/abs/1911.10206}{{\ttfamily arXiv:1911.10206 [hep-ph]}}.

\bibitem{Chao:2019smr}
W.~Chao and Y.~Liu, ``{CP violation in the top-assisted electroweak baryogenesis},'' \href{https://arxiv.org/abs/1910.09303}{{\ttfamily arXiv:1910.09303 [hep-ph]}}.

\bibitem{Xie:2020wzn}
K.-P. Xie, ``{Lepton-mediated electroweak baryogenesis, gravitational waves and the $4\tau$ final state at the collider},'' \href{https://dx.doi.org/10.1007/JHEP02(2021)090}{{\em JHEP} {\bfseries 02} (2021) 090}, \href{https://arxiv.org/abs/2011.04821}{{\ttfamily arXiv:2011.04821 [hep-ph]}}. [Erratum: JHEP 8, 052 (2022)].

\bibitem{Xie:2020bkl}
K.-P. Xie, L.~Bian, and Y.~Wu, ``{Electroweak baryogenesis and gravitational waves in a composite Higgs model with high dimensional fermion representations},'' \href{https://dx.doi.org/10.1007/JHEP12(2020)047}{{\em JHEP} {\bfseries 12} (2020) 047}, \href{https://arxiv.org/abs/2005.13552}{{\ttfamily arXiv:2005.13552 [hep-ph]}}.

\bibitem{Fernandez-Martinez:2020szk}
E.~Fern\'andez-Mart\'\i{}nez, J.~L\'opez-Pav\'on, T.~Ota, and S.~Rosauro-Alcaraz, ``{$\nu$ electroweak baryogenesis},'' \href{https://dx.doi.org/10.1007/JHEP10(2020)063}{{\em JHEP} {\bfseries 10} (2020) 063}, \href{https://arxiv.org/abs/2007.11008}{{\ttfamily arXiv:2007.11008 [hep-ph]}}.

\bibitem{Basler:2021kgq}
P.~Basler, L.~Biermann, M.~M\"uhlleitner, and J.~M\"uller, ``{Electroweak baryogenesis in the CP-violating two-Higgs doublet model},'' \href{https://dx.doi.org/10.1140/epjc/s10052-023-11192-9}{{\em Eur. Phys. J. C} {\bfseries 83} no.~1, (2023) 57}, \href{https://arxiv.org/abs/2108.03580}{{\ttfamily arXiv:2108.03580 [hep-ph]}}.

\bibitem{Biermann:2022meg}
L.~Biermann, M.~M{\"u}hlleitner, and J.~M{\"u}ller, ``{Electroweak phase transition in a dark sector with CP violation},'' \href{https://dx.doi.org/10.1140/epjc/s10052-023-11612-w}{{\em Eur. Phys. J. C} {\bfseries 83} no.~5, (2023) 439}, \href{https://arxiv.org/abs/2204.13425}{{\ttfamily arXiv:2204.13425 [hep-ph]}}.

\bibitem{Anisha:2022hgv}
Anisha, L.~Biermann, C.~Englert, and M.~M\"uhlleitner, ``{Two Higgs doublets, effective interactions and a strong first-order electroweak phase transition},'' \href{https://dx.doi.org/10.1007/JHEP08(2022)091}{{\em JHEP} {\bfseries 08} (2022) 091}, \href{https://arxiv.org/abs/2204.06966}{{\ttfamily arXiv:2204.06966 [hep-ph]}}.

\bibitem{Anisha:2023vvu}
Anisha, D.~Azevedo, L.~Biermann, C.~Englert, and M.~M\"uhlleitner, ``{Effective 2HDM Yukawa interactions and a strong first-order electroweak phase transition},'' \href{https://dx.doi.org/10.1007/JHEP02(2024)045}{{\em JHEP} {\bfseries 02} (2024) 045}, \href{https://arxiv.org/abs/2311.06353}{{\ttfamily arXiv:2311.06353 [hep-ph]}}.

\bibitem{Fernandez-Martinez:2022stj}
E.~Fern\'andez-Mart\'\i{}nez, J.~L\'opez-Pav\'on, J.~M. No, T.~Ota, and S.~Rosauro-Alcaraz, ``{$\nu $ Electroweak baryogenesis: the scalar singlet strikes back},'' \href{https://dx.doi.org/10.1140/epjc/s10052-023-11887-z}{{\em Eur. Phys. J. C} {\bfseries 83} no.~8, (2023) 715}, \href{https://arxiv.org/abs/2210.16279}{{\ttfamily arXiv:2210.16279 [hep-ph]}}.

\bibitem{Harigaya:2023bmp}
K.~Harigaya and I.~R. Wang, ``{ALP-assisted strong first-order electroweak phase transition and baryogenesis},'' \href{https://dx.doi.org/10.1007/JHEP04(2024)108}{{\em JHEP} {\bfseries 04} (2024) 108}, \href{https://arxiv.org/abs/2309.00587}{{\ttfamily arXiv:2309.00587 [hep-ph]}}.

\bibitem{Carena:2021onl}
M.~Carena, C.~Krause, Z.~Liu, and Y.~Wang, ``{New approach to electroweak symmetry nonrestoration},'' \href{https://dx.doi.org/10.1103/PhysRevD.104.055016}{{\em Phys. Rev. D} {\bfseries 104} no.~5, (2021) 055016}, \href{https://arxiv.org/abs/2104.00638}{{\ttfamily arXiv:2104.00638 [hep-ph]}}.

\bibitem{Chang:2022psj}
J.~H. Chang, M.~O. Olea-Romacho, and E.~H. Tanin, ``{Electroweak asymmetric early Universe via a scalar condensate},'' \href{https://dx.doi.org/10.1103/PhysRevD.106.113003}{{\em Phys. Rev. D} {\bfseries 106} no.~11, (2022) 113003}, \href{https://arxiv.org/abs/2210.05680}{{\ttfamily arXiv:2210.05680 [hep-ph]}}.

\bibitem{Kuzmin:1985mm}
V.~A. Kuzmin, V.~A. Rubakov, and M.~E. Shaposhnikov, ``{On the Anomalous Electroweak Baryon Number Nonconservation in the Early Universe},'' \href{https://dx.doi.org/10.1016/0370-2693(85)91028-7}{{\em Phys. Lett. B} {\bfseries 155} (1985) 36}.

\bibitem{Shaposhnikov:1986jp}
M.~E. Shaposhnikov, ``{Possible Appearance of the Baryon Asymmetry of the Universe in an Electroweak Theory},'' {\em JETP Lett.} {\bfseries 44} (1986) 465--468.

\bibitem{Farrar:1993hn}
G.~R. Farrar and M.~E. Shaposhnikov, ``{Baryon asymmetry of the universe in the standard electroweak theory},'' \href{https://dx.doi.org/10.1103/PhysRevD.50.774}{{\em Phys. Rev. D} {\bfseries 50} (1994) 774}, \href{https://arxiv.org/abs/hep-ph/9305275}{{\ttfamily arXiv:hep-ph/9305275}}.

\bibitem{Farrar:1993sp}
G.~R. Farrar and M.~E. Shaposhnikov, ``{Baryon asymmetry of the universe in the minimal Standard Model},'' \href{https://dx.doi.org/10.1103/PhysRevLett.70.2833}{{\em Phys. Rev. Lett.} {\bfseries 70} (1993) 2833--2836}, \href{https://arxiv.org/abs/hep-ph/9305274}{{\ttfamily arXiv:hep-ph/9305274}}. [Erratum: Phys.Rev.Lett. 71, 210 (1993)].

\bibitem{Kamionkowski:1993fg}
M.~Kamionkowski, A.~Kosowsky, and M.~S. Turner, ``{Gravitational radiation from first order phase transitions},'' \href{https://dx.doi.org/10.1103/PhysRevD.49.2837}{{\em Phys. Rev. D} {\bfseries 49} (1994) 2837--2851}, \href{https://arxiv.org/abs/astro-ph/9310044}{{\ttfamily arXiv:astro-ph/9310044}}.

\bibitem{Caprini:2007xq}
C.~Caprini, R.~Durrer, and G.~Servant, ``{Gravitational wave generation from bubble collisions in first-order phase transitions: An analytic approach},'' \href{https://dx.doi.org/10.1103/PhysRevD.77.124015}{{\em Phys. Rev. D} {\bfseries 77} (2008) 124015}, \href{https://arxiv.org/abs/0711.2593}{{\ttfamily arXiv:0711.2593 [astro-ph]}}.

\bibitem{Huber:2008hg}
S.~J. Huber and T.~Konstandin, ``{Gravitational Wave Production by Collisions: More Bubbles},'' \href{https://dx.doi.org/10.1088/1475-7516/2008/09/022}{{\em JCAP} {\bfseries 09} (2008) 022}, \href{https://arxiv.org/abs/0806.1828}{{\ttfamily arXiv:0806.1828 [hep-ph]}}.

\bibitem{Hindmarsh:2013xza}
M.~Hindmarsh, S.~J. Huber, K.~Rummukainen, and D.~J. Weir, ``{Gravitational waves from the sound of a first order phase transition},'' \href{https://dx.doi.org/10.1103/PhysRevLett.112.041301}{{\em Phys. Rev. Lett.} {\bfseries 112} (2014) 041301}, \href{https://arxiv.org/abs/1304.2433}{{\ttfamily arXiv:1304.2433 [hep-ph]}}.

\bibitem{Hindmarsh:2016lnk}
M.~Hindmarsh, ``{Sound shell model for acoustic gravitational wave production at a first-order phase transition in the early Universe},'' \href{https://dx.doi.org/10.1103/PhysRevLett.120.071301}{{\em Phys. Rev. Lett.} {\bfseries 120} no.~7, (2018) 071301}, \href{https://arxiv.org/abs/1608.04735}{{\ttfamily arXiv:1608.04735 [astro-ph.CO]}}.

\bibitem{Kosowsky:2002}
A.~Kosowsky, A.~Mack, and T.~Kahniashvili, ``Gravitational radiation from cosmological turbulence,'' \href{https://dx.doi.org/10.1103/PhysRevD.66.024030}{{\em Phys. Rev. D} {\bfseries 66} (Jul, 2002) 024030}. \url{https://link.aps.org/doi/10.1103/PhysRevD.66.024030}.

\bibitem{Caprini:2006}
C.~Caprini and R.~Durrer, ``Gravitational waves from stochastic relativistic sources: Primordial turbulence and magnetic fields,'' \href{https://dx.doi.org/10.1103/PhysRevD.74.063521}{{\em Phys. Rev. D} {\bfseries 74} (Sep, 2006) 063521}. \url{https://link.aps.org/doi/10.1103/PhysRevD.74.063521}.

\bibitem{Caprini:2009yp}
C.~Caprini, R.~Durrer, and G.~Servant, ``{The stochastic gravitational wave background from turbulence and magnetic fields generated by a first-order phase transition},'' \href{https://dx.doi.org/10.1088/1475-7516/2009/12/024}{{\em JCAP} {\bfseries 12} (2009) 024}, \href{https://arxiv.org/abs/0909.0622}{{\ttfamily arXiv:0909.0622 [astro-ph.CO]}}.

\bibitem{Wainwright:2011kj}
C.~L. Wainwright, ``{CosmoTransitions: Computing Cosmological Phase Transition Temperatures and Bubble Profiles with Multiple Fields},'' \href{https://dx.doi.org/10.1016/j.cpc.2012.04.004}{{\em Comput. Phys. Commun.} {\bfseries 183} (2012) 2006--2013}, \href{https://arxiv.org/abs/1109.4189}{{\ttfamily arXiv:1109.4189 [hep-ph]}}.

\bibitem{Menzo:2025cim}
T.~Menzo, A.~Roman, S.~Gleyzer, K.~Matchev, G.~T. Fleming, S.~H{\"o}che, S.~Mrenna, and P.~Shyamsundar, ``{HEPTAPOD: Orchestrating High Energy Physics Workflows Towards Autonomous Agency},'' \href{https://arxiv.org/abs/2512.15867}{{\ttfamily arXiv:2512.15867 [hep-ph]}}.

\bibitem{Agrawal:2026lvg}
P.~Agrawal, N.~Craig, A.~Madden, and I.~V. Lombera, ``{The FERMIACC: Agents for Particle Theory},'' \href{https://arxiv.org/abs/2603.22538}{{\ttfamily arXiv:2603.22538 [hep-ph]}}.

\bibitem{Desai:2026nmx}
A.~Desai, ``{RooAgent: An LLM Agent for Root-Based High Energy Physics Analysis},'' \href{https://arxiv.org/abs/2605.17318}{{\ttfamily arXiv:2605.17318 [hep-ph]}}.

\bibitem{Menzo:2026qrl}
T.~Menzo, A.~Roman, G.~T. Fleming, S.~Gleyzer, K.~T. Matchev, and S.~Mrenna, ``{Agentic Diagrammatica: Towards Autonomous Symbolic Computation in High Energy Physics},'' \href{https://arxiv.org/abs/2603.26990}{{\ttfamily arXiv:2603.26990 [hep-ph]}}.

\bibitem{Qiu:2026iby}
S.~Qiu, Z.~Cai, J.~Wei, Z.~Li, Y.~Yin, Q.-H. Cao, C.~Liu, M.-x. Luo, X.-B. Yuan, and H.~X. Zhu, ``{An End-to-end Architecture for Collider Physics and Beyond},'' \href{https://arxiv.org/abs/2603.14553}{{\ttfamily arXiv:2603.14553 [hep-ph]}}.

\bibitem{Bakshi:2025fgx}
S.~D. Bakshi {\em et~al.}, ``{ArgoLOOM: agentic AI for fundamental physics from quarks to cosmos},'' \href{https://arxiv.org/abs/2510.02426}{{\ttfamily arXiv:2510.02426 [hep-ph]}}.

\bibitem{Plehn:2026gxv}
T.~Plehn, D.~Schiller, and N.~Schmal, ``{MadAgents},'' \href{https://arxiv.org/abs/2601.21015}{{\ttfamily arXiv:2601.21015 [hep-ph]}}.

\bibitem{Faroughy:2026dkj}
D.~A. Faroughy, S.~Palacios~Schweitzer, I.~Pang, S.~Mishra-Sharma, and D.~Shih, ``{Collider-Bench: Benchmarking AI Agents with Particle Physics Analysis Reproduction},'' \href{https://arxiv.org/abs/2605.13950}{{\ttfamily arXiv:2605.13950 [cs.LG]}}.

\bibitem{Lucente:2026kgh}
M.~Lucente, S.~Pascoli, F.~Sala, and M.~Zandi, ``{DarkAgents},'' \href{https://arxiv.org/abs/2606.11157}{{\ttfamily arXiv:2606.11157 [hep-ph]}}.

\bibitem{Gent:2025csq}
T.~Gent, S.~Huber, K.~Mimasu, and J.~M. No, ``{Towards precise baryogenesis in the 2HDM$+a$},'' \href{https://arxiv.org/abs/2512.22081}{{\ttfamily arXiv:2512.22081 [hep-ph]}}.

\bibitem{Parwani:1991gq}
R.~R. Parwani, ``{Resummation in a hot scalar field theory},'' \href{https://dx.doi.org/10.1103/PhysRevD.45.4695}{{\em Phys. Rev. D} {\bfseries 45} (1992) 4695}, \href{https://arxiv.org/abs/hep-ph/9204216}{{\ttfamily arXiv:hep-ph/9204216}}. [Erratum: Phys.Rev.D 48, 5965 (1993)].

\bibitem{Friedlander:2020tnq}
A.~Friedlander, I.~Banta, J.~M. Cline, and D.~Tucker-Smith, ``{Wall speed and shape in singlet-assisted strong electroweak phase transitions},'' \href{https://dx.doi.org/10.1103/PhysRevD.103.055020}{{\em Phys. Rev. D} {\bfseries 103} no.~5, (2021) 055020}, \href{https://arxiv.org/abs/2009.14295}{{\ttfamily arXiv:2009.14295 [hep-ph]}}.

\bibitem{Arnold:1992rz}
P.~B. Arnold and O.~Espinosa, ``{The Effective potential and first order phase transitions: Beyond leading-order},'' \href{https://dx.doi.org/10.1103/PhysRevD.47.3546}{{\em Phys. Rev. D} {\bfseries 47} (1993) 3546}, \href{https://arxiv.org/abs/hep-ph/9212235}{{\ttfamily arXiv:hep-ph/9212235}}. [Erratum: Phys.Rev.D 50, 6662 (1994)].

\bibitem{Laurent:2022jrs}
B.~Laurent and J.~M. Cline, ``{First principles determination of bubble wall velocity},'' \href{https://dx.doi.org/10.1103/PhysRevD.106.023501}{{\em Phys. Rev. D} {\bfseries 106} no.~2, (2022) 023501}, \href{https://arxiv.org/abs/2204.13120}{{\ttfamily arXiv:2204.13120 [hep-ph]}}.

\bibitem{DeCurtis:2022hlx}
S.~De~Curtis, L.~D. Rose, A.~Guiggiani, A.~G. Muyor, and G.~Panico, ``{Bubble wall dynamics at the electroweak phase transition},'' \href{https://dx.doi.org/10.1007/JHEP03(2022)163}{{\em JHEP} {\bfseries 03} (2022) 163}, \href{https://arxiv.org/abs/2201.08220}{{\ttfamily arXiv:2201.08220 [hep-ph]}}.

\bibitem{DeCurtis:2023hil}
S.~De~Curtis, L.~Delle~Rose, A.~Guiggiani, A.~Gil~Muyor, and G.~Panico, ``{Collision integrals for cosmological phase transitions},'' \href{https://dx.doi.org/10.1007/JHEP05(2023)194}{{\em JHEP} {\bfseries 05} (2023) 194}, \href{https://arxiv.org/abs/2303.05846}{{\ttfamily arXiv:2303.05846 [hep-ph]}}.

\bibitem{DeCurtis:2024hvh}
S.~De~Curtis, L.~Delle~Rose, A.~Guiggiani, A.~Gil~Muyor, and G.~Panico, ``{Non-linearities in cosmological bubble wall dynamics},'' \href{https://dx.doi.org/10.1007/JHEP05(2024)009}{{\em JHEP} {\bfseries 05} (2024) 009}, \href{https://arxiv.org/abs/2401.13522}{{\ttfamily arXiv:2401.13522 [hep-ph]}}.

\bibitem{Ai:2021kak}
W.-Y. Ai, B.~Garbrecht, and C.~Tamarit, ``{Bubble wall velocities in local equilibrium},'' \href{https://dx.doi.org/10.1088/1475-7516/2022/03/015}{{\em JCAP} {\bfseries 03} no.~03, (2022) 015}, \href{https://arxiv.org/abs/2109.13710}{{\ttfamily arXiv:2109.13710 [hep-ph]}}.

\bibitem{Ai:2023see}
W.-Y. Ai, B.~Laurent, and J.~van~de Vis, ``{Model-independent bubble wall velocities in local thermal equilibrium},'' \href{https://dx.doi.org/10.1088/1475-7516/2023/07/002}{{\em JCAP} {\bfseries 07} (2023) 002}, \href{https://arxiv.org/abs/2303.10171}{{\ttfamily arXiv:2303.10171 [astro-ph.CO]}}.

\bibitem{Ai:2024shx}
W.-Y. Ai, X.~Nagels, and M.~Vanvlasselaer, ``{Criterion for ultra-fast bubble walls: the impact of hydrodynamic obstruction},'' \href{https://dx.doi.org/10.1088/1475-7516/2024/03/037}{{\em JCAP} {\bfseries 03} (2024) 037}, \href{https://arxiv.org/abs/2401.05911}{{\ttfamily arXiv:2401.05911 [hep-ph]}}.

\bibitem{Ramsey-Musolf:2025jyk}
M.~J. Ramsey-Musolf and J.~Zhu, ``{Bubble wall velocity from Kadanoff-Baym equations: fluid dynamics and microscopic interactions},'' \href{https://arxiv.org/abs/2504.13724}{{\ttfamily arXiv:2504.13724 [hep-ph]}}.

\bibitem{Ai:2025bjw}
W.-Y. Ai, M.~Carosi, B.~Garbrecht, C.~Tamarit, and M.~Vanvlasselaer, ``{Bubble wall dynamics from nonequilibrium quantum field theory},'' \href{https://dx.doi.org/10.1007/JHEP08(2025)077}{{\em JHEP} {\bfseries 08} (2025) 077}, \href{https://arxiv.org/abs/2504.13725}{{\ttfamily arXiv:2504.13725 [hep-ph]}}.

\bibitem{Ai:2024btx}
W.-Y. Ai, B.~Laurent, and J.~van~de Vis, ``{Bounds on the bubble wall velocity},'' \href{https://dx.doi.org/10.1007/JHEP02(2025)119}{{\em JHEP} {\bfseries 02} (2025) 119}, \href{https://arxiv.org/abs/2411.13641}{{\ttfamily arXiv:2411.13641 [hep-ph]}}.

\bibitem{Carena:2025flp}
M.~Carena, A.~Ireland, T.~Ou, and I.~R. Wang, ``{The discriminant power of bubble wall velocities: gravitational waves and electroweak baryogenesis},'' \href{https://dx.doi.org/10.1007/JHEP09(2025)175}{{\em JHEP} {\bfseries 09} (2025) 175}, \href{https://arxiv.org/abs/2504.17841}{{\ttfamily arXiv:2504.17841 [hep-ph]}}.

\bibitem{Krajewski:2024gma}
T.~Krajewski, M.~Lewicki, and M.~Zych, ``{Bubble-wall velocity in local thermal equilibrium: hydrodynamical simulations vs analytical treatment},'' \href{https://dx.doi.org/10.1007/JHEP05(2024)011}{{\em JHEP} {\bfseries 05} (2024) 011}, \href{https://arxiv.org/abs/2402.15408}{{\ttfamily arXiv:2402.15408 [astro-ph.CO]}}.

\bibitem{Krajewski:2024zxg}
T.~Krajewski, M.~Lewicki, I.~Nalecz, and M.~Zych, ``{Steady-state bubbles beyond local thermal equilibrium},'' \href{https://dx.doi.org/10.1007/JHEP06(2025)118}{{\em JHEP} {\bfseries 06} (2025) 118}, \href{https://arxiv.org/abs/2411.16580}{{\ttfamily arXiv:2411.16580 [astro-ph.CO]}}.

\bibitem{Eriksson:2025owh}
M.~Eriksson and M.~Laine, ``{Entropy production at electroweak bubble walls from scalar field fluctuations},'' \href{https://dx.doi.org/10.1088/1475-7516/2025/09/027}{{\em JCAP} {\bfseries 09} (2025) 027}, \href{https://arxiv.org/abs/2507.07755}{{\ttfamily arXiv:2507.07755 [hep-ph]}}.

\bibitem{Ginsparg:1980ef}
P.~H. Ginsparg, ``{First Order and Second Order Phase Transitions in Gauge Theories at Finite Temperature},'' \href{https://dx.doi.org/10.1016/0550-3213(80)90418-6}{{\em Nucl. Phys. B} {\bfseries 170} (1980) 388--408}.

\bibitem{Appelquist:1981vg}
T.~Appelquist and R.~D. Pisarski, ``{High-Temperature Yang-Mills Theories and Three-Dimensional Quantum Chromodynamics},'' \href{https://dx.doi.org/10.1103/PhysRevD.23.2305}{{\em Phys. Rev. D} {\bfseries 23} (1981) 2305}.

\bibitem{Braaten:1995cm}
E.~Braaten and A.~Nieto, ``{Effective field theory approach to high temperature thermodynamics},'' \href{https://dx.doi.org/10.1103/PhysRevD.51.6990}{{\em Phys. Rev. D} {\bfseries 51} (1995) 6990--7006}, \href{https://arxiv.org/abs/hep-ph/9501375}{{\ttfamily arXiv:hep-ph/9501375}}.

\bibitem{Kajantie:1995dw}
K.~Kajantie, M.~Laine, K.~Rummukainen, and M.~E. Shaposhnikov, ``{Generic rules for high temperature dimensional reduction and their application to the standard model},'' \href{https://dx.doi.org/10.1016/0550-3213(95)00549-8}{{\em Nucl. Phys. B} {\bfseries 458} (1996) 90--136}, \href{https://arxiv.org/abs/hep-ph/9508379}{{\ttfamily arXiv:hep-ph/9508379}}.

\bibitem{Croon:2020cgk}
D.~Croon, O.~Gould, P.~Schicho, T.~V.~I. Tenkanen, and G.~White, ``{Theoretical uncertainties for cosmological first-order phase transitions},'' \href{https://dx.doi.org/10.1007/JHEP04(2021)055}{{\em JHEP} {\bfseries 04} (2021) 055}, \href{https://arxiv.org/abs/2009.10080}{{\ttfamily arXiv:2009.10080 [hep-ph]}}.

\bibitem{Gould:2021dzl}
O.~Gould, ``{Real scalar phase transitions: a nonperturbative analysis},'' \href{https://dx.doi.org/10.1007/JHEP04(2021)057}{{\em JHEP} {\bfseries 04} (2021) 057}, \href{https://arxiv.org/abs/2101.05528}{{\ttfamily arXiv:2101.05528 [hep-ph]}}.

\bibitem{Niemi:2021qvp}
L.~Niemi, P.~Schicho, and T.~V.~I. Tenkanen, ``{Singlet-assisted electroweak phase transition at two loops},'' \href{https://dx.doi.org/10.1103/PhysRevD.103.115035}{{\em Phys. Rev. D} {\bfseries 103} no.~11, (2021) 115035}, \href{https://arxiv.org/abs/2103.07467}{{\ttfamily arXiv:2103.07467 [hep-ph]}}.

\bibitem{Schicho:2021gca}
P.~M. Schicho, T.~V.~I. Tenkanen, and J.~\"Osterman, ``{Robust approach to thermal resummation: Standard Model meets a singlet},'' \href{https://dx.doi.org/10.1007/JHEP06(2021)130}{{\em JHEP} {\bfseries 06} (2021) 130}, \href{https://arxiv.org/abs/2102.11145}{{\ttfamily arXiv:2102.11145 [hep-ph]}}.

\bibitem{Ekstedt:2022bff}
A.~Ekstedt, P.~Schicho, and T.~V.~I. Tenkanen, ``{DRalgo: A package for effective field theory approach for thermal phase transitions},'' \href{https://dx.doi.org/10.1016/j.cpc.2023.108725}{{\em Comput. Phys. Commun.} {\bfseries 288} (2023) 108725}, \href{https://arxiv.org/abs/2205.08815}{{\ttfamily arXiv:2205.08815 [hep-ph]}}.

\bibitem{Brdar:2019fur}
V.~Brdar, L.~Graf, A.~J. Helmboldt, and X.-J. Xu, ``{Gravitational Waves as a Probe of Left-Right Symmetry Breaking},'' \href{https://dx.doi.org/10.1088/1475-7516/2019/12/027}{{\em JCAP} {\bfseries 12} (2019) 027}, \href{https://arxiv.org/abs/1909.02018}{{\ttfamily arXiv:1909.02018 [hep-ph]}}.

\bibitem{Harigaya:2022wzt}
K.~Harigaya and I.~R. Wang, ``{Baryogenesis in a parity solution to the strong CP problem},'' \href{https://dx.doi.org/10.1007/JHEP11(2023)189}{{\em JHEP} {\bfseries 11} (2023) 189}, \href{https://arxiv.org/abs/2210.16207}{{\ttfamily arXiv:2210.16207 [hep-ph]}}.

\bibitem{Hall:2019ank}
E.~Hall, T.~Konstandin, R.~McGehee, H.~Murayama, and G.~Servant, ``{Baryogenesis From a Dark First-Order Phase Transition},'' \href{https://dx.doi.org/10.1007/JHEP04(2020)042}{{\em JHEP} {\bfseries 04} (2020) 042}, \href{https://arxiv.org/abs/1910.08068}{{\ttfamily arXiv:1910.08068 [hep-ph]}}.

\bibitem{Hall:2019rld}
E.~Hall, T.~Konstandin, R.~McGehee, and H.~Murayama, ``{Asymmetric matter from a dark first-order phase transition},'' \href{https://dx.doi.org/10.1103/PhysRevD.107.055011}{{\em Phys. Rev. D} {\bfseries 107} no.~5, (2023) 055011}, \href{https://arxiv.org/abs/1911.12342}{{\ttfamily arXiv:1911.12342 [hep-ph]}}.

\bibitem{Goncalves:2025uwh}
J.~Gon{\c{c}}alves, D.~Marfatia, A.~P. Morais, and R.~Pasechnik, ``{Supercooled phase transitions in conformal dark sectors explain NANOGrav data},'' \href{https://dx.doi.org/10.1016/j.physletb.2025.139829}{{\em Phys. Lett. B} {\bfseries 869} (2025) 139829}, \href{https://arxiv.org/abs/2501.11619}{{\ttfamily arXiv:2501.11619 [hep-ph]}}.

\bibitem{Balan:2025uke}
S.~Balan, T.~Bringmann, F.~Kahlhoefer, J.~Matuszak, and C.~Tasillo, ``{Sub-GeV dark matter and nano-Hertz gravitational waves from a classically conformal dark sector},'' \href{https://dx.doi.org/10.1088/1475-7516/2025/08/062}{{\em JCAP} {\bfseries 08} (2025) 062}, \href{https://arxiv.org/abs/2502.19478}{{\ttfamily arXiv:2502.19478 [hep-ph]}}.

\bibitem{Costa:2025csj}
F.~Costa, J.~Hoefken~Zink, M.~Lucente, S.~Pascoli, and S.~Rosauro-Alcaraz, ``{Supercooled dark scalar phase transitions explanation of NANOGrav data},'' \href{https://dx.doi.org/10.1016/j.physletb.2025.139634}{{\em Phys. Lett. B} {\bfseries 868} (2025) 139634}, \href{https://arxiv.org/abs/2501.15649}{{\ttfamily arXiv:2501.15649 [hep-ph]}}.

\bibitem{Kannike:2012pe}
K.~Kannike, ``{Vacuum Stability Conditions From Copositivity Criteria},'' \href{https://dx.doi.org/10.1140/epjc/s10052-012-2093-z}{{\em Eur. Phys. J. C} {\bfseries 72} (2012) 2093}, \href{https://arxiv.org/abs/1205.3781}{{\ttfamily arXiv:1205.3781 [hep-ph]}}.

\bibitem{ParticleDataGroup:2024cfk}
{\bfseries Particle Data Group} Collaboration, S.~Navas {\em et~al.}, ``{Review of particle physics},'' \href{https://dx.doi.org/10.1103/PhysRevD.110.030001}{{\em Phys. Rev. D} {\bfseries 110} no.~3, (2024) 030001}.

\bibitem{Cline:2011mm}
J.~M. Cline, K.~Kainulainen, and M.~Trott, ``{Electroweak Baryogenesis in Two Higgs Doublet Models and B meson anomalies},'' \href{https://dx.doi.org/10.1007/JHEP11(2011)089}{{\em JHEP} {\bfseries 11} (2011) 089}, \href{https://arxiv.org/abs/1107.3559}{{\ttfamily arXiv:1107.3559 [hep-ph]}}.

\bibitem{Carena:2022qpf}
M.~Carena, Y.-Y. Li, T.~Ou, and Y.~Wang, ``{Anatomy of the electroweak phase transition for dark sector induced baryogenesis},'' \href{https://dx.doi.org/10.1007/JHEP02(2023)139}{{\em JHEP} {\bfseries 02} (2023) 139}, \href{https://arxiv.org/abs/2210.14352}{{\ttfamily arXiv:2210.14352 [hep-ph]}}.

\bibitem{Bittar:2025lcr}
P.~Bittar, S.~Roy, and C.~E.~M. Wagner, ``{Self consistent thermal resummation: a case study of the phase transition in 2HDM},'' \href{https://dx.doi.org/10.1007/JHEP12(2025)021}{{\em JHEP} {\bfseries 12} (2025) 021}, \href{https://arxiv.org/abs/2504.02024}{{\ttfamily arXiv:2504.02024 [hep-ph]}}.

\bibitem{Bahl:2024ykv}
H.~Bahl, M.~Carena, A.~Ireland, and C.~E.~M. Wagner, ``{Improved thermal resummation for multi-field potentials},'' \href{https://dx.doi.org/10.1007/JHEP09(2024)153}{{\em JHEP} {\bfseries 09} (2024) 153}, \href{https://arxiv.org/abs/2404.12439}{{\ttfamily arXiv:2404.12439 [hep-ph]}}.

\bibitem{Ekstedt:2020abj}
A.~Ekstedt and J.~L{\"o}fgren, ``{A Critical Look at the Electroweak Phase Transition},'' \href{https://dx.doi.org/10.1007/JHEP12(2020)136}{{\em JHEP} {\bfseries 12} (2020) 136}, \href{https://arxiv.org/abs/2006.12614}{{\ttfamily arXiv:2006.12614 [hep-ph]}}.

\bibitem{Curtin:2016urg}
D.~Curtin, P.~Meade, and H.~Ramani, ``{Thermal Resummation and Phase Transitions},'' \href{https://dx.doi.org/10.1140/epjc/s10052-018-6268-0}{{\em Eur. Phys. J. C} {\bfseries 78} no.~9, (2018) 787}, \href{https://arxiv.org/abs/1612.00466}{{\ttfamily arXiv:1612.00466 [hep-ph]}}.

\bibitem{Boyd:1993tz}
C.~G. Boyd, D.~E. Brahm, and S.~D.~H. Hsu, ``{Resummation methods at finite temperature: The Tadpole way},'' \href{https://dx.doi.org/10.1103/PhysRevD.48.4963}{{\em Phys. Rev. D} {\bfseries 48} (1993) 4963--4973}, \href{https://arxiv.org/abs/hep-ph/9304254}{{\ttfamily arXiv:hep-ph/9304254}}.

\bibitem{Hindmarsh:2015qta}
M.~Hindmarsh, S.~J. Huber, K.~Rummukainen, and D.~J. Weir, ``{Numerical simulations of acoustically generated gravitational waves at a first order phase transition},'' \href{https://dx.doi.org/10.1103/PhysRevD.92.123009}{{\em Phys. Rev. D} {\bfseries 92} no.~12, (2015) 123009}, \href{https://arxiv.org/abs/1504.03291}{{\ttfamily arXiv:1504.03291 [astro-ph.CO]}}.

\bibitem{Caprini:2015zlo}
C.~Caprini {\em et~al.}, ``{Science with the space-based interferometer eLISA. II: Gravitational waves from cosmological phase transitions},'' \href{https://dx.doi.org/10.1088/1475-7516/2016/04/001}{{\em JCAP} {\bfseries 04} (2016) 001}, \href{https://arxiv.org/abs/1512.06239}{{\ttfamily arXiv:1512.06239 [astro-ph.CO]}}.

\bibitem{Hindmarsh:2017gnf}
M.~Hindmarsh, S.~J. Huber, K.~Rummukainen, and D.~J. Weir, ``{Shape of the acoustic gravitational wave power spectrum from a first order phase transition},'' \href{https://dx.doi.org/10.1103/PhysRevD.96.103520}{{\em Phys. Rev. D} {\bfseries 96} no.~10, (2017) 103520}, \href{https://arxiv.org/abs/1704.05871}{{\ttfamily arXiv:1704.05871 [astro-ph.CO]}}. [Erratum: Phys.Rev.D 101, 089902 (2020)].

\bibitem{Ellis:2020awk}
J.~Ellis, M.~Lewicki, and J.~M. No, ``{Gravitational waves from first-order cosmological phase transitions: lifetime of the sound wave source},'' \href{https://dx.doi.org/10.1088/1475-7516/2020/07/050}{{\em JCAP} {\bfseries 07} (2020) 050}, \href{https://arxiv.org/abs/2003.07360}{{\ttfamily arXiv:2003.07360 [hep-ph]}}.

\bibitem{Caprini:2024hue}
{\bfseries LISA Cosmology Working Group} Collaboration, C.~Caprini, R.~Jinno, M.~Lewicki, E.~Madge, M.~Merchand, G.~Nardini, M.~Pieroni, A.~Roper~Pol, and V.~Vaskonen, ``{Gravitational waves from first-order phase transitions in LISA: reconstruction pipeline and physics interpretation},'' \href{https://dx.doi.org/10.1088/1475-7516/2024/10/020}{{\em JCAP} {\bfseries 10} (2024) 020}, \href{https://arxiv.org/abs/2403.03723}{{\ttfamily arXiv:2403.03723 [astro-ph.CO]}}.

\bibitem{Espinosa:2010hh}
J.~R. Espinosa, T.~Konstandin, J.~M. No, and G.~Servant, ``{Energy Budget of Cosmological First-order Phase Transitions},'' \href{https://dx.doi.org/10.1088/1475-7516/2010/06/028}{{\em JCAP} {\bfseries 06} (2010) 028}, \href{https://arxiv.org/abs/1004.4187}{{\ttfamily arXiv:1004.4187 [hep-ph]}}.

\bibitem{Azevedo:2018fmj}
D.~Azevedo, P.~M. Ferreira, M.~M. Muhlleitner, S.~Patel, R.~Santos, and J.~Wittbrodt, ``{CP in the dark},'' \href{https://dx.doi.org/10.1007/JHEP11(2018)091}{{\em JHEP} {\bfseries 11} (2018) 091}, \href{https://arxiv.org/abs/1807.10322}{{\ttfamily arXiv:1807.10322 [hep-ph]}}.

\end{thebibliography}\endgroup

\clearpage
\newpage
\onecolumngrid

\setcounter{section}{0}
\setcounter{subsection}{0}
\setcounter{subsubsection}{0}
\setcounter{equation}{0}
\setcounter{figure}{0}
\setcounter{table}{0}

\makeatletter
\renewcommand{\thesection}{S\arabic{section}}
\renewcommand{\thesubsection}{\thesection.\arabic{subsection}}
\renewcommand{\thesubsubsection}{\thesubsection.\arabic{subsubsection}}

\renewcommand{\theequation}{S\arabic{equation}}
\renewcommand{\thefigure}{S\arabic{figure}}
\renewcommand{\thetable}{S\arabic{table}}

\renewcommand{\theHsection}{S\arabic{section}}
\renewcommand{\theHsubsection}{S\arabic{section}.\arabic{subsection}}
\renewcommand{\theHsubsubsection}{S\arabic{section}.\arabic{subsection}.\arabic{subsubsection}}
\renewcommand{\theHequation}{S\arabic{equation}}
\renewcommand{\theHfigure}{S\arabic{figure}}
\renewcommand{\theHtable}{S\arabic{table}}
\makeatother

\begin{center}
    \textbf{\large \texttt{LeWRON}: Agentic Analysis of Electroweak Phase Transitions} \\ 
    \vspace{0.5cm}
    { \it \large Supplemental Material}\\ 
    \vspace{0.5cm}
    {Isaac R. Wang}
\end{center}

Here we provide supplemental material for the main text.
In Sec.~\ref{sec:physics}, we summarize the physics framework employed in the computation pipeline of \texttt{LeWRON}.
In Sec.~\ref{sec:install}, we briefly describe how to install and run the agent. Detailed instructions can be found on \href{https://github.com/quarkquartet/LeWRON}{GitHub}.
In Sec.~\ref{sec:model}, we provide details of the BSM models in our discovery mode examples.

\section{Physics scheme of \texttt{LeWRON}}
\label{sec:physics}

This section reviews the main physics ingredients used by \texttt{LeWRON}.

\subsection{$T=0$ physics}

The first step in an EWPT calculation is to identify the real scalar directions on which the effective potential will be constructed. \texttt{LeWRON} starts from the primitive component fields in the Lagrangian and treats every real scalar degree of freedom as a candidate background direction, unless one of the following elimination criteria applies:
\begin{itemize}
\item \textbf{Gauge-fixing of Goldstone directions.} Would-be Goldstone modes associated with broken gauge generators are removed from the background field space. In a basis where a Goldstone boson is a VEV-weighted mixture of several primitive components, only the Goldstone subspace is eliminated; the orthogonal combinations are physical scalar directions and are retained unless another criterion removes them.

\item \textbf{Exact symmetry.} A field direction is eliminated when the full Lagrangian is invariant under a symmetry that fixes its expectation value to zero, and all retained candidate vacuum sectors preserve that symmetry. An exact symmetry of the Lagrangian is therefore not, by itself, sufficient to eliminate a field if the phase-transition path is allowed to break it spontaneously.

\item \textbf{Charge-conservation ansatz.} Charged or color-breaking directions are set to zero by default when the target EWPT analysis is restricted to electromagnetic- and color-preserving backgrounds. For example, charge-breaking VEVs in a 2HDM are eliminated unless the user explicitly requests a charge-breaking sector.
\end{itemize}
All remaining neutral physical directions are retained as background candidates.

After identifying the relevant degrees of freedom for the phase transition, \texttt{LeWRON} derives the tree-level quantities needed for the analysis, including the mass spectrum and parametrization relations.
The scalar mass spectrum is derived by computing
\begin{align}
  \mathcal{M}_{ij} = \frac{\partial^2 V_0}{\partial \phi_i \phi_j}
\end{align}
for all degrees of freedom $\phi_i$, $\phi_j$.
The gauge-boson mass spectrum is derived from the standard formula
\begin{align}
  \mathcal{M}_{\text{gauge},ij} = \sum_i \frac{\partial^2 (D_\mu \Phi)^2}{\partial A_\mu^i \partial A_\mu^j}\,,
\end{align}
where $A_\mu^i$ denotes the electroweak gauge boson degrees of freedom, and $\Phi_i$ denotes any $SU(2)_L$ doublet scalar. The fermion mass spectrum can be read directly from the Yukawa interactions.
For analytically derivable relations between Lagrangian-level parameters and physical observables, specifically when the mixing matrix is at most 2-by-2, \texttt{LeWRON} computes translation formulae that replace Lagrangian parameters with physical observables.

Three constraints are then derived at the tree level: the bounded-from-below
condition, following Ref.~\cite{Kannike:2012pe}, the requirement
that the electroweak vacuum be the global minimum, and the non-tachyonic
requirement at the electroweak vacuum.
The bounded-from-below condition is imposed on the quartic part of the full
gauge-invariant scalar potential,
\begin{equation}
    V_4(\Phi_i)>0
    \qquad
    \text{for all gauge-inequivalent directions at large field values}.
\end{equation}
Equivalently, after rewriting \(V_4\) in terms of non-negative gauge-invariant
radial variables and minimizing over any bounded internal orbit variables, the
resulting form is required to be copositive.  This step is performed on the full
scalar potential rather than on the reduced neutral background potential used
for the finite-temperature analysis.
The electroweak vacuum is then required to be the global tree-level minimum,
\begin{equation}
    V_0(\phi_{\rm EW}) < V_0(\phi_\alpha),
\end{equation}
for every inequivalent stationary point \(\phi_\alpha\) of \(V_0\).  This removes
parameter points for which the desired electroweak vacuum is only metastable
already at zero temperature.
At tree level, some analytic conditions can be derived, while \texttt{LeWRON} implements a full numerical check of the global-minimum condition at loop level.
Finally, the scalar Hessian at the electroweak vacuum is required to have no
tachyonic physical eigenvalues,
\begin{equation}
    \left(M_S^2\right)_{ij}
    =
    \left.
    \frac{\partial^2 V_0}{\partial \phi_i \partial \phi_j}
    \right|_{\phi=\phi_{\rm EW}},
    \qquad
    m_{\rm phys}^2>0 .
\end{equation}
Would-be Goldstone directions are treated separately: after the tadpole
conditions are imposed, their eigenvalues must vanish exactly, as required by
Goldstone's theorem, and are not converted into positivity constraints.

The zero-temperature one-loop correction is implemented through the Coleman--Weinberg potential in the $\overline{\rm MS}$ scheme,
\begin{align}
   V_{\mathrm{CW}}(\phi_i)
   = \frac{1}{64\pi^2}
   \sum_i n_i \hat m_i^4(\phi_i)
   \left[
      \ln\left(\frac{\hat m_i^2(\phi_i)}{\mu_R^2}\right) - c_i
   \right] ,
\end{align}
where $\hat m_i^2(\phi_i)$ are the field-dependent mass eigenvalues and $n_i$ denotes the corresponding number of degrees of freedom, including the fermionic sign.
We take $c_i=3/2$ for scalars, fermions, and longitudinal gauge bosons, and $c_i=1/2$ for transverse gauge bosons.
By default, \texttt{LeWRON} chooses the renormalization scale $\mu_R=m_Z=91.1880~{\rm GeV}$.
The remaining input parameters, including gauge and Yukawa couplings and particle masses, are taken from the 2024 Particle Data Group values renormalized at $m_Z$~\cite{ParticleDataGroup:2024cfk}.
Goldstone infrared (IR) divergences can arise at the vev when derivatives of $V_{\rm CW}$ are evaluated.
\texttt{LeWRON} employs the widely used IR-cutoff method, in which vanishing Goldstone boson masses are replaced by the Higgs boson mass~\cite{Cline:2011mm}.

\texttt{LeWRON} supports two standard treatments of one-loop boundary conditions.
The first directly solves for the input parameters from the one-loop vacuum and mass conditions, as done for example in Refs.~\cite{Carena:2019une,Harigaya:2023bmp}.
The second introduces a counterterm potential \(V_{\rm CT}\) that preserves the tree-level vacuum expectation values, physical masses, and mixing angles after \(V_{\rm CW}\) is included; see, e.g., Refs.~\cite{Friedlander:2020tnq,Carena:2022qpf,Gent:2025csq}.
In the counterterm prescription, \texttt{LeWRON} constructs
\begin{align}
    V_{\rm CT}(\phi)
    =
    \sum_a \delta p_a
    \frac{\partial V_0(\phi)}{\partial p_a}
    + \sum_i \delta T_i \phi_i ,
    \label{eq:counterterm_general}
\end{align}
where \(p_a\) are the parameters of the tree-level potential and \(\phi_i\) denotes the retained nonzero background fields.
The first term corresponds to finite shifts of the tree-level parameters.
The second term contains optional tadpole counterterms, introduced only when parameter variations alone cannot cancel a required one-loop tadpole condition.

The finite coefficients \(\{\delta p_a,\delta T_i\}\) are fixed by requiring that the one-loop correction does not shift the chosen \(T=0\) vacuum or the tree-level scalar spectrum,
\begin{align}
    \left.
    \frac{\partial}{\partial \phi_i}
    \left(V_{\rm CW}+V_{\rm CT}\right)
    \right|_{\phi=\langle\phi\rangle}
    &=0 ,
    \label{eq:ct_tadpole}
    \\
    \left.
    \frac{\partial^2}{\partial \phi_i \partial \phi_j}
    \left(V_{\rm CW}+V_{\rm CT}\right)
    \right|_{\phi=\langle\phi\rangle}
    &=0 ,
    \label{eq:ct_hessian}
\end{align}
for the nontrivial tadpole and independent Hessian conditions in the retained background-field basis.
Symmetry-protected vanishing tadpoles are omitted, while accidental zeros at \(\phi=\langle\phi\rangle\) are retained as renormalization conditions.
If a parameter is invisible to the first- and second-derivative conditions, \texttt{LeWRON} adds the lowest-order derivative condition at which that parameter first appears,
\begin{align}
    \left.
    \frac{\partial^n}{\partial\phi_{i_1}\cdots\partial\phi_{i_n}}
    \left(V_{\rm CW}+V_{\rm CT}\right)
    \right|_{\phi=\langle\phi\rangle}
    =0 ,
    \qquad n>2 .
    \label{eq:ct_higher_derivative}
\end{align}
All selected conditions are then assembled into a linear system,
\begin{align}
    M\,\delta \mathbf{q} = - \mathbf{b},
    \qquad
    \delta\mathbf{q}
    =
    \left(\delta p_1,\delta p_2,\ldots,\delta T_1,\delta T_2,\ldots\right)^T ,
    \label{eq:ct_linear_system}
\end{align}
where \(\mathbf b\) contains the corresponding Coleman--Weinberg derivatives evaluated at \(\phi=\langle\phi\rangle\).
Rank-deficient systems are repaired, when possible, by adding the minimal tadpole counterterm that contributes to the deficient condition.

The counterterm prescription is used by default.
In reproduction mode, or upon user request, \texttt{LeWRON} instead follows the renormalization prescription specified by the input reference or by the user.

\subsection{Finite-$T$ physics}

\texttt{LeWRON} supports the traditional one-loop computation of the thermal effective potential
\begin{equation}\label{eq:VT}
    V_{T}^{1\text{-loop}}(\phi_i, T) = \frac{T^4}{2\pi^2} \sum_i n_i J_{B/F} \left( \frac{\hat{m}_i(\phi_i)^2}{T^2} \right) \,, 
\end{equation}
where again $i$ runs over all particle species.
The bosonic and fermionic thermal functions are defined as
\begin{equation}
    J_{B/F}(y^2) = \int_0^\infty dx\, x^2 \ln \left( 1 \mp e^{- \sqrt{x^2 + y^2}} \right) \,.
\end{equation}
\texttt{LeWRON} uses \verb|CosmoTransitions| as the backend to numerically implement the thermal potential.
This public package, however, does not handle the imaginary part of this integral properly when the squared mass is negative.
We use the modified version published in \href{https://gitlab.com/claudius-krause/ew_nr}{GitLab}, which was used to implement the model in Ref.~\cite{Carena:2021onl}.

Thermal resummation must be treated consistently. By default, \texttt{LeWRON} uses Parwani resummation~\cite{Parwani:1991gq}, which replaces the boson masses by their thermal masses throughout the one-loop potential.
Arnold-Espinosa resummation~\cite{Arnold:1992rz}, another widely implemented alternative, is included as an optional choice.
However, we note that Arnold-Espinosa resummation is known to be problematic at high temperature and can sometimes lead to unexpected thermal histories, including non-restoration (see Refs.~\cite{Bittar:2025lcr,Bahl:2024ykv} for recent case studies).
As complementary options, we provide two additional choices. First, improved Parwani resummation manually subtracts the two-loop pieces that Parwani resummation implicitly includes in the CW logarithms~\cite{Ekstedt:2020abj},
\begin{align}
    V_{\mathrm{EL}}
  = V_{\mathrm{Parwani}}
  - T^2 \sum_{i\,\in\,\text{bosons}} c_i^2\,
    \left.\frac{\partial V_{\mathrm{CW}}}{\partial m_i^2}\right|_{m_i^2\,\text{unresummed}},
\end{align}
Second, partial-dressing resummation~\cite{Curtin:2016urg,Curtin:2016urg,Boyd:1993tz,Bahl:2024ykv,Bittar:2025lcr} solves self-consistently for $\bar M^2_i$ at each $(\phi, T)$:
\begin{align}
    \bar M^2_i \;=\; m_i^2(\phi) + \Pi_i\!\left(\bar M^2, T\right)
\end{align}
and then re-evaluates $\Pi_i$ on the right-hand side using the resummed $\bar M^2$
at each iteration.

After the finite-temperature effective potential is evaluated, we compute the critical temperature, where the true and false vacua are degenerate, and the nucleation temperature, where the bubble nucleation rate per Hubble volume reaches unity.
The criterion for the latter is taken to be $S_{3}/T \simeq 140$, where $S_3$ is the 3-dimensional tunneling action.

In the context of electroweak phase transitions, gravitational waves are sourced by sound waves and MHD turbulence.
The GW spectrum from the sound wave takes the form~\cite{Hindmarsh:2013xza,Hindmarsh:2015qta,Caprini:2015zlo,Hindmarsh:2016lnk,Hindmarsh:2017gnf,Ellis:2020awk,Caprini:2024hue}
\begin{align}
    \Omega_{\rm sw}h^2  = 2.65 \times 10^{-6} \left( \frac{\kappa_{\rm sw} \, \alpha_n}{1 + \alpha_n} \right)^2 \left( \frac{H}{\beta} \right) \left( \frac{100}{g_*} \right)^{1/3} (H \tau_{\rm sw}) \, v_w \left( \frac{f}{f_{\rm sw}} \right)^3 \left( \frac{7}{4 + 3(f/f_{\rm sw})^2} \right)^{7/2} \,,
\end{align}
the sound wave peak frequency $f_{\rm sw}$ set by
\begin{align}
    \label{eq:fsw}
    f_{\rm sw} = 1.9 \times 10^{-5} \, \mathrm{Hz} \left( \frac{1}{v_w} \right) \left( \frac{\beta}{H} \right) \left( \frac{g_*(T_n)}{100} \right)^{1/6} \left( \frac{T_n}{100~\rm GeV} \right) \,.
\end{align}
The energy fraction that goes into the sound wave, $\kappa_{\rm sw}$, is computed from the numerical fits in Appendix A of Ref.~\cite{Espinosa:2010hh}.
The MHD turbulence spectrum is not thoroughly understood.
We use the widely adopted result from modeling Kolmogorov-type turbulence~\cite{Caprini:2009yp,Caprini:2015zlo,Hindmarsh:2013xza,Hindmarsh:2015qta,Hindmarsh:2016lnk,Hindmarsh:2017gnf}
\begin{align}
    \Omega_{\rm tur}h^2 = 3.35 \times 10^{-4} \left( \frac{\kappa_{\rm turb} \, \alpha_n}{1 + \alpha_n} \right)^{3/2} \left( \frac{H}{\beta} \right) \left( \frac{100}{g_*} \right)^{1/3} v_w S_{\rm turb} \,,
\end{align}
where
\begin{align}
    \label{eq:turb}
     & S_{\rm turb} = \frac{(f/f_{\rm turb})^3}{(1 + (f/f_{\rm turb}))^{11/3} (1+8\pi f/h_*)} \,, \nonumber                                                                                                        \\
     & f_{\rm turb} = 2.7 \times 10^{-5} \, \mathrm{Hz} \left( \frac{1}{v_w} \right) \left( \frac{\beta}{H} \right) \left( \frac{g_*(T_n)}{100} \right)^{1/6} \left( \frac{T_n}{100~\rm GeV} \right) \,, \nonumber \\
     & h_* = 16.5 \times 10^{-6} \, \mathrm{Hz} \left( \frac{g_*(T_n)}{100} \right)^{1/6} \left( \frac{T_n}{100~\rm GeV} \right) \,.
\end{align}
The coefficient $\kappa_{\rm tur}$ describes the fraction of vacuum energy that is transformed into MHD turbulence, and is estimated to be between $(0.05-0.1)\kappa_{\rm sw}$ for thermal phase transitions.

\section{Installation and example usage}
\label{sec:install}

\texttt{LeWRON} is distributed as a \verb|Python| package with a console
command \verb|lewron|.  A minimal installation is
\begin{commandbox}
git clone https://github.com/quarkquartet/LeWRON
cd LeWRON
conda create -n lewron python=3.11
conda activate lewron
pip install .
\end{commandbox}
The toolbox stages use the Anthropic API,
\begin{commandbox}
export ANTHROPIC_API_KEY=sk-ant-...
\end{commandbox}
while the interactive Explorer uses the local \verb|claude| command-line login.
The numerical \verb|CosmoTransitions| backend is installed by the agent when it
is first needed and the selected version is recorded in the run state.

A discovery-mode run starts from a model-description file and writes all
generated artifacts, code, notes, and outputs to a run directory,
\begin{commandbox}
lewron run --input examples/ALP.md --run-dir examples/alp
\end{commandbox}
whereas a reproduction-mode run can start from an inline literature target,
\begin{commandbox}
lewron run --input "Reproduce Table 1 of 2009.14295, neglecting the wall velocity."
\end{commandbox}
An interrupted or completed run is resumed by specifying only the run directory,
\begin{commandbox}
lewron run --run-dir examples/alp
\end{commandbox}

We use the ALP-Higgs model as an example to illustrate the basic workflow in the terminal interface.
We first write the model in a Markdown file with the following content:
\begin{commandbox}
The model extends the SM by a light pseudo-Nambu-Goldstone boson (ALP) $S$, arising from the spontaneous breaking of a global $\mathrm{U}(1)$ symmetry at a scale $f \gg v$
The tree-level potential is:
$$
V(H,S) = - (\mu_H^2  - A f \cos \delta)|H|^2 + \lambda |H|^4 + \mu_S^2 f^2 \left(1 - \cos (\frac{S}{f})\right) - A f \left(|H|^2 - v^2\right) \cos (\frac{S}{f} - \delta),
$$
where $H$ is the SM Higgs doublet. The ALP mass $m_S \sim \mu_S$ lies in the MeV-GeV range, far below the electroweak scale. The Higgs parameters are fixed by $m_h = 125.20\,\text{GeV}$ and $v \simeq 174\,\text{GeV}$.

At $T = 0$, the Higgs acquires its standard VEV and $\langle S\rangle = 0$.
We are interested in the case where $0 < \delta < \pi/2$.
The field value of $S$ is considered within one period, that is, the period containing 0. Other periods are beyond the scope.

We are interested in a strongly first-order electroweak phase transition.
\end{commandbox}

Then, after typing the following command, the starting interface appears:
\begin{commandbox}
+--------------------------------------------------------------+
|        _            __        __ ____    ___   _   _         |
|       | |      ___  \ \      / /|  _ \  / _ \ | \ | |        |
|       | |     / _ \  \ \ /\ / / | |_) || | | ||  \| |        |
|       | |___ |  __/   \ V  V /  |  _ < | |_| || |\  |        |
|       |_____| \___|    \_/\_/   |_| \_\ \___/ |_| \_|        |
|                                                              |
|                            * * *                             |
|                         *         *                          |
|                        *  ⟨H⟩ > 0  *                         |
|                         *         *                          |
|                            * * *                             |
+--------------------------------------------------------------+
|  version  1.0.0      https://github.com/quarkquartet/LeWRON  |
|  cite     arXiv:XXXX.XXXXX  (placeholder)                    |
+--------------------------------------------------------------+
Input:   examples/ALP.md
Run dir: examples/alp
Mode:      discovery
Build:     full
Request:   Analyze the strongly first-order electroweak phase transition in the SM extended by an ALP scalar S with the given cosine potential, scanning over the model parameters (mu_S, A, f, delta) in the regime 0 < delta < pi/2.
Project:   alp
Analyzing model symmetries...
\end{commandbox}
where the last sentence is dynamically updated according to the running stage of the agent.

At the end of stage 1a, it prints a compact summary for the user to inspect:
\begin{commandbox}
symmetry analysis

Scalar fields:
  - $H$ [H] -- $SU(2)_L$ doublet, $Y=1/2$
  - $S$ [S] -- Real gauge singlet (ALP),
    $SU(2)_L\times U(1)_Y$ singlet

Background directions:
  - $h$ [h] (from h)
  - $S$ [S_field] (from S_field)

  [warning] $S$: The phase $\delta\in(0,\pi/2)$ in
  $\cos(S/f-\delta)$ explicitly breaks the $S\to-S$ symmetry, so
  $S$ cannot be eliminated by a symmetry argument and must remain
  a background direction.

Eliminated directions:
  - $\mathrm{Re}\,\phi^+$ [phi_re]  [charge_conservation]
  - $\mathrm{Im}\,\phi^+$ [phi_im]  [charge_conservation]
  - $G^0$ [G0]  [gauge_fixing]

  Charge-breaking directions (phi_re, phi_im) set to zero by
  assumption; verify no charge-breaking minimum downstream.

Candidate vacua:
  - true_vev (active: h)
  - EW-symmetric sector (active: S_field)
  - Coupled (h, S) sector (active: h, S_field)
\end{commandbox}

It also prints reminders for the user to check:
\begin{commandbox}
   The symmetry analysis stage is complete. Review the proposal:                                         
   examples/alp/notes/report.md 
   examples/alp/artifacts/model_spec.json
   Press Enter to accept, or type a comment - ask a question, dispute the result, or tell me what to change.
   For a multi-line comment, press Option+Enter or Ctrl+J to start a new line; Enter sends.
\end{commandbox}

The user can enter a message here to revise the output, after which the agent reruns the stage and replies.
After confirmation, stage 1b starts running, and the user sees the following information:
\begin{commandbox}
Background fields: h, S_field
Scalar mass blocks: 4
- [phi_re] → Gplus_re (diagonal)
- [phi_im] → Gplus_im (diagonal)
- [h, S_field] → h_1, h_2 (analytic_2x2) 
- [G0] → G0 (diagonal)                   
Gauge bosons: 3                            
- W (ndof 6)
- Z (ndof 3)
- photon (ndof 2)
Fermions: 1
- top (ndof 12)
Parametrization (analytic):
INPUT params (physical observables):     3
DERIVED params (Lagrangian eliminated):  3
REMAINING params (free Lagrangian):      4
total Lagrangian = derived + remaining = 7; total benchmark inputs = input + remaining = 7
physical inputs (substituted for Lagrangian params):
- m_h_2
- m_S_field_2
- theta_h_S_field
Lagrangian parameters eliminated:
- mu_H2
- lam
- mu_S2
remaining free Lagrangian parameters (specify per benchmark):
- A
- f
- delta
- v
Full report: notes/report.md
Artifact: artifacts/model_spec.json
\end{commandbox}
Here, the agent gives an incorrect parametrization result: the mixing angle is not used to solve for $A$.
This happens because the input Lagrangian is ambiguous. The vev $v$ and mass term $\mu_H^2$ are not independent variables, but both appear in the Lagrangian, so the agent counts the wrong number of parameters.
The user can enter a revision directly at this checkpoint, e.g.,
\begin{commandbox}
    The parametrization needs to be revised. $v$ should be physical inputs, $A$ should be eliminated, and remaining is $f$ and $\delta$. So there are total 6 parameters in this model, 4 solved (2 masses, $v$ and mixing angle) and 2 remaining.
\end{commandbox}
Stages 2 and 3 then follow a similar running pattern.

After tools are built up, the agent starts the Explorer module by confirming the task:
\begin{commandbox}
Explorer

Explorer ready -- a Claude coding agent working in this run's directory.
It reads the model, writes its own analysis code, and reports results.
It starts in plan mode (it proposes first). Type /edit to let it work --
it applies file edits automatically and asks before each shell command.
/bypass is off -- re-launch with `lewron run --allow-bypass` (or set
LEWRON_ALLOW_BYPASS=1) to enable it. The input line below is always live --
type any time (Option+Enter or Ctrl+J for a new line; Enter sends).
Commands: /plan, /edit, /bypass, /mode <name>, /model <name>, /status,
/btw <message>, /stop, /quit, /help. Typing a message while it works stops
the current step and steers it; use /btw to send a note (e.g. a status
question) without stopping it.

Task for this run

Analyze the strongly first-order EWPT in an SM + ALP
(pseudo-Nambu-Goldstone) model with cosine portal potential, computing the
one-loop effective potential with Parwani resummation and bubble nucleation
temperature T_n defined by S_3(T_n)/T_n = 140, with 1-loop parametrization
fixed by physical Higgs observables.
\end{commandbox}

The user then directly interact with the Explorer module in the same terminal window, whose backend is Claude Code SDK, with our customized experience library installed.
The user can also choose to quit the program and run the generated code with Claude Code directly.
\texttt{LeWRON} provides options to install the experience library as Claude Code skill.
Users are also provided interface to install or write their own experience.

\section{Further details for example models}
\label{sec:model}

Here, we provide more details on the models for the two discovery mode examples.
The two reproduction-mode examples can be found in the corresponding references.

\subsection{The ALP-Higgs model}
For the ALP-Higgs discovery-mode example, the scalar sector contains the SM
Higgs doublet \(H\) and a real ALP field \(S\).
We follow the convention of Ref.~\cite{Harigaya:2023bmp} and write the potential as
\begin{align}
V_0(H,S) ={}&
-\left(\mu_H^2-Af\cos\delta\right)H^\dagger H
+\lambda\left(H^\dagger H\right)^2
+\mu_S^2f^2\left(1-\cos\frac{S}{f}\right)
-Af\left(H^\dagger H-v^2\right)
\cos\left(\frac{S}{f}-\delta\right).
\label{eq:alp_v0}
\end{align}
Here \(v\) is defined by \(\langle H^\dagger H\rangle=v^2\).  After
eliminating the Goldstone directions, we write the neutral background as
\(H=(0,h/\sqrt{2})^T\), so the electroweak vacuum is
\((h,S)=(\sqrt{2}v,0)\).  The tree-level stationarity condition gives
\begin{align}
    \mu_H^2=2\lambda v^2 .
\end{align}
The key feature of this model is that the effective Higgs quartic coupling is reduced by $Av/\mu_S^2$, so even a light ALP can enhance the EWPT~\cite{Jeong:2018jqe,Jeong:2018ucz,Harigaya:2023bmp}.
The field-value shift of $S$ during a phase transition is estimated as
\begin{align}
    \Delta S \propto \frac{A v^2 }{\mu_S^2}\sin \delta\,,
\end{align}
and we parameterize the decay constant as $f \equiv c A v^2 \sin \delta/\mu_S^2$, with $c$ as the input parameter.
$\Delta S$ is enhanced for a light ALP mass, which can lead to a large baryon asymmetry if an appropriate CP-violating source is present.

After reducing to the neutral background, the counterterm potential is written
with one augmented linear counterterm for the ALP tadpole:
\begin{align}
V_{\rm CT}(h,S) ={}&
-\frac{\delta\mu_H^2}{2}h^2
+\frac{\delta\lambda}{4}h^4
+\delta\mu_S^2 f^2\left(1-\cos\frac{S}{f}\right)
\nonumber\\
&+\delta A\left[
\frac{fh^2\cos\delta}{2}
-\frac{fh^2}{2}\cos\left(\frac{S}{f}-\delta\right)
+fv^2\cos\left(\frac{S}{f}-\delta\right)
\right]
+t_S S .
\label{eq:alp_vct}
\end{align}
The renormalization conditions are imposed after reducing to the neutral
background \((h,S)\), at \((h,S)=(\sqrt{2}v,0)\).  We denote the
Coleman--Weinberg tadpoles by \(T_i^{\rm CW}\) and the second derivatives by
\(A_{ij}^{\rm CW}\).  The counterterms
\(\delta\mu_H^2\), \(\delta\lambda\), \(\delta\mu_S^2\), \(\delta A\), and
\(t_S\) are fixed by
\begin{align}
\sqrt{2}v\left(-\delta\mu_H^2+2\delta\lambda v^2\right)
&= -T_h^{\rm CW},
\nonumber\\
t_S
&= -T_S^{\rm CW},
\nonumber\\
-\delta\mu_H^2+6\delta\lambda v^2
&= -A_{hh}^{\rm CW},
\nonumber\\
-\sqrt{2}v\sin\delta\,\delta A
&= -A_{hS}^{\rm CW},
\nonumber\\
\delta\mu_S^2
&= -A_{SS}^{\rm CW}.
\label{eq:alp_ct_conditions}
\end{align}

\subsection{The 2HDM+$S$ model}
For the \(2\mathrm{HDM}+S\) discovery-mode example, the finite-temperature
analysis starts from two \(SU(2)_L\) doublets \(\Phi_1,\Phi_2\) and a real
singlet \(\Phi_S\), where a $\mathbb{Z}_2$ symmetry is imposed on $\Phi_2$ and $S$.
The tree-level potential is~\cite{Biermann:2022meg,Azevedo:2018fmj}
\begin{align}\label{eq:scalarpotential}
    V =&  m_{11}^2 |\Phi_1|^2 + \frac{1}{2} \lambda_1 |\Phi_1|^4 + m_{22}^2 |\Phi_2|^2 + \frac{1}{2} \lambda_2 |\Phi_2|^4 + \frac{1}{2}m_S^2 S^2 + \frac{1}{4}\lambda_S S^4 \nonumber \\
    &  + (A \Phi_1^\dagger \Phi_2 S + \mathrm{h.c.}) \nonumber \\
    & + \lambda_3 |\Phi_1|^2 |\Phi_2|^2+ \lambda_4 |\Phi_1^\dagger \Phi_2|^2 + \frac{1}{2} \left(\lambda_5(\Phi_1^\dagger \Phi_2)^2 + \mathrm{h.c.}\right) \nonumber \\
    & + \frac{1}{2} \lambda_{1S} |\Phi_1|^2 S^2 + \frac{1}{2} \lambda_{2S} |\Phi_2|^2 S^2\,.
\end{align}
In this model, the only complex parameter is $A \equiv A_{\rm re} + i A_{\rm im}$.
We write the doublet fields as
\begin{align}
    \Phi_1 = \begin{pmatrix}
        G^+ \\
        \frac{1}{\sqrt{2}}(\rho_1 + i G_0)
    \end{pmatrix}\,,~ \Phi_2 = \begin{pmatrix}
        H^+\\
        \frac{1}{\sqrt{2}}(\rho_2 + i \eta_2) 
    \end{pmatrix}
    \label{eq:2hdoublets}
\end{align}
We are interested in the parameter space where the EW vacuum is
\(\langle\Phi_1^0\rangle=v/\sqrt{2}\), \(\langle\Phi_2\rangle=0\), and
\(\langle\Phi_S\rangle=0\), giving
\begin{align}
    m_{11}^2=-\frac{\lambda_1 v^2}{2}.
\end{align}
This isolates the CP-violating phase in the $\Phi_2$-$S$ sector, thereby avoiding electric-dipole-moment constraints.

We identify the background field degrees of freedom as $(\rho_1, \rho_2, \eta_2, S)$.
The counterterm potential used in the on-shell tree-level-mass scheme is
\begin{align}
V_{\rm CT} ={}&
\frac{\delta m_{11}^2}{2}\rho_1^2
+ \frac{\delta m_{22}^2}{2}(\rho_2^2+\eta_2^2)
+ \frac{\delta m_S^2}{2}S^2
+ \frac{\delta\lambda_1}{8}\rho_1^4
+ \frac{\delta\lambda_2}{8}(\rho_2^2+\eta_2^2)^2
\nonumber\\
&+ \frac{\delta\lambda_3}{4}\rho_1^2(\rho_2^2+\eta_2^2)
+ \frac{\delta\lambda_4}{4}\rho_1^2(\rho_2^2+\eta_2^2)
+ \frac{\delta\lambda_5}{4}\rho_1^2(\rho_2^2-\eta_2^2)
\nonumber\\
&+ \frac{\delta\lambda_6}{4}S^4
+ \frac{\delta\lambda_7}{4}S^2\rho_1^2
+ \frac{\delta\lambda_8}{4}S^2(\rho_2^2+\eta_2^2)
+ \delta A_{\rm re}S\rho_1\rho_2
- \delta A_{\rm im}S\eta_2\rho_1 .
\label{eq:2hdms_vct}
\end{align}
All counterterm conditions below are evaluated at
\((\rho_1,\rho_2,\eta_2,S)=(v,0,0,0)\).  We denote the Coleman--Weinberg
tadpole by \(T_i^{\rm CW}\), the second derivative by \(A_{ij}^{\rm CW}\),
and the fourth derivative by \(B_{ijkl}^{\rm CW}\).  The conditions generated
by \texttt{LeWRON} are
\begin{align}
\delta m_{11}^2 v + \frac{1}{2}\delta\lambda_1 v^3
&= -T_{\rho_1}^{\rm CW},
\nonumber\\
\delta m_{11}^2 + \frac{3}{2}\delta\lambda_1 v^2
&= -A_{\rho_1\rho_1}^{\rm CW},
\nonumber\\
\delta m_{22}^2
+ \frac{v^2}{2}(\delta\lambda_3+\delta\lambda_4+\delta\lambda_5)
&= -A_{\rho_2\rho_2}^{\rm CW},
\nonumber\\
\delta m_{22}^2
+ \frac{v^2}{2}(\delta\lambda_3+\delta\lambda_4-\delta\lambda_5)
&= -A_{\eta_2\eta_2}^{\rm CW},
\nonumber\\
\delta m_S^2 + \frac{1}{2}\delta\lambda_7 v^2
&= -A_{SS}^{\rm CW},
\nonumber\\
\delta A_{\rm re}v
&= -A_{\rho_2 S}^{\rm CW},
\nonumber\\
-\delta A_{\rm im}v
&= -A_{\eta_2 S}^{\rm CW},
\nonumber\\
\delta m_{22}^2 + \frac{1}{2}\delta\lambda_3 v^2
&= -A_{H^\pm H^\pm}^{\rm CW},
\nonumber\\
\delta\lambda_3+\delta\lambda_4+\delta\lambda_5
&= -B_{\rho_1\rho_1\rho_2\rho_2}^{\rm CW},
\nonumber\\
3\delta\lambda_2
&= -B_{\rho_2\rho_2\rho_2\rho_2}^{\rm CW},
\nonumber\\
6\delta\lambda_6
&= -B_{SSSS}^{\rm CW},
\nonumber\\
\delta\lambda_7
&= -B_{\rho_1\rho_1SS}^{\rm CW},
\nonumber\\
\delta\lambda_8
&= -B_{\rho_2\rho_2SS}^{\rm CW}.
\label{eq:2hdms_ct_conditions}
\end{align}
The \(\mathbb{Z}_2\) symmetry protects the tadpoles for
\(\rho_2\), \(\eta_2\), and \(S\), so no separate counterterm tadpole
conditions are imposed for these directions.
\end{document}